\documentclass{jfm}

\usepackage{graphicx}
\usepackage{natbib}
\usepackage{subcaption}
\usepackage{color}
\ifCUPmtlplainloaded \else
  \checkfont{eurm10}
  \iffontfound
    \IfFileExists{upmath.sty}
      {\typeout{^^JFound AMS Euler Roman fonts on the system,
                   using the 'upmath' package.^^J}%
       \usepackage{upmath}}
      {\typeout{^^JFound AMS Euler Roman fonts on the system, but you
                   dont seem to have the}%
       \typeout{'upmath' package installed. JFM.cls can take advantage
                 of these fonts,^^Jif you use 'upmath' package.^^J}%
      }
  \else
  \fi
\fi

\ifCUPmtlplainloaded \else
  \checkfont{msam10}
  \iffontfound
    \IfFileExists{amssymb.sty}
      {\typeout{^^JFound AMS Symbol fonts on the system, using the
                'amssymb' package.^^J}%
       \usepackage{amssymb}%

      }{}
  \fi
\fi

\ifCUPmtlplainloaded \else
  \IfFileExists{amsbsy.sty}
    {\typeout{^^JFound the 'amsbsy' package on the system, using it.^^J}%
     \usepackage{amsbsy}}
    {\providecommand\boldsymbol[1]{\mbox{\boldmath $##1$}}}
\fi

\newsavebox{\astrutbox}
\sbox{\astrutbox}{\rule[-5pt]{0pt}{20pt}}

\title[Two-phase Taylor Couette: Euler-Lagrange simulations]{Drag reduction in numerical two-phase Taylor-Couette turbulence using an Euler-Lagrange approach}

\author[Vamsi Spandan, Rodolfo Ostilla-M{\'o}nico, Roberto Verzicco, Detlef Lohse]%
{Vamsi Spandan$^1$%
  ,\ns
Rodolfo Ostilla-M{\'o}nico$^1$,\ns Roberto Verzicco$^{2,1}$, and Detlef Lohse$^{1,3}$%
\thanks{Email address for correspondence: d.lohse@utwente.nl }}

\affiliation{$^1$Physics of Fluids Group, Faculty of Science and Technology, J.M. Burgers Center for Fluid Dynamics and MESA+ Institute, University of Twente, 7500 AE Enschede, Netherlands\\[\affilskip]
$^2$Dipartimento di Ingegneria Meccanica,University of Rome \lq Tor Vergata\rq, Via del Politecnico 1,
Rome 00133, Italy\\[\affilskip]
$^3$Max Planck Institute for Dynamics and Self-Organization, 37077, G\"ottingen, Germany
}

\pubyear{?}
\volume{?}
\pagerange{?}
\date{?; revised ?; accepted ?. - To be entered by editorial office}

\begin{document}

\maketitle

\begin{abstract}

Two-phase turbulent Taylor-Couette (TC) flow is simulated using an Euler-Lagrange approach to study the effects of a secondary phase dispersed into a turbulent carrier phase (here bubbles dispersed into water). The dynamics of the carrier phase is computed using Direct Numerical Simulations (DNS) in an Eulerian framework, while the bubbles are tracked in a Lagrangian manner by modelling the effective drag, lift, added mass and buoyancy force acting on them. Two-way coupling is implemented between the dispersed phase and the carrier phase which allows for momentum exchange among both phases and to study the effect of the dispersed phase on the carrier phase dynamics. The radius ratio of the TC setup is fixed to $\eta$=0.833, and a maximum inner cylinder Reynolds number of $Re_i$=8000 is reached. We vary the Froude number ($Fr$), which is the ratio of the centripetal to the gravitational acceleration of the dispersed phase and study its effect on the net torque required to drive the TC system. 

For the two-phase TC system, we observe drag reduction, i.e., the torque required to drive the inner cylinder is less compared to that of the single phase system. The net drag reduction decreases with increasing Reynolds number $Re_i$, which is consistent with previous experimental findings \citep{murai2005bubble,murai2008frictional}. The drag reduction is strongly related to the Froude number: for fixed Reynolds number we observe higher drag reduction when $Fr<1$ than for with $Fr>1$. This buoyancy effect is more prominent in low $Re_i$ systems and decreases with increasing Reynolds number $Re_i$. We trace the drag reduction back to the weakening of the angular momentum carrying Taylor rolls by the rising bubbles. We also investigate how the motion of the dispersed phase depends on $Re_i$ and $Fr$, by studying the individual trajectories and mean dispersion of bubbles in the radial and axial directions. Indeed, the less buoyant bubbles (large $Fr$) tend to get trapped by the Taylor rolls, while the more buoyant bubbles (small Fr) rise through and weaken them.     

\end{abstract}

\begin{keywords}
Taylor-Couette flow, drag reduction, bubbly flow, Euler-Lagrange scheme
\end{keywords}

\section{Introduction}

Frictional losses in the form of drag in turbulent flows is a major drain of energy in applications related to process technology, naval transportation, and transport of liquified natural gas in pipelines \citep{ceccio2010friction}. It has been known for a long time that the injection of a small concentration of a dispersed phase into a carrier fluid can result in significant drag reduction, making it of interest for fundamental scientific research in order to understand the mechanism and optimise the effect for engineering applications. Drag reduction has been demonstrated in many physical systems in the past, like bubbles injected into a turbulent boundary layer over a flat plate \citep{madavan1983reduction, madavan1985measurements, xu2002numerical}, bubbles in channel flows \citep{lu2005effect,murai2007skin,lu2008effect,dabiri2013transition,pang2014numerical,park2015drag}, and Taylor-Couette (TC) flow \citep{murai2005bubble, murai2008frictional, van2005drag, van2007bubbly, van2013importance}; the reader is referred to \citet{ceccio2010friction} and \citet{murai2014frictional} for more detailed reviews. The magnitude of reduction in drag or the driving force for a two-phase system when compared to a single phase system can be massive; \citet{van2013importance} showed drag reduction of up to 40 \% with just 4 \% of bubbles dispersed into TC flow. Various theories have been suggested to explain the origin of this effect; among them are theories based on effective compressibility \citep{ferrante2004physical,l2005drag}, for low Reynolds number flows disruption of coherent vortical structures present in the single phase flow \citep{sugiyama2008microbubbly}, and also effects of bubble deformability \citep{van2007bubbly,van2013importance}. However, the exact mechanism is still unknown and it maybe expected that different mechanisms such as those mentioned above maybe dominant in different flow regimes. 

TC flow has been one of the classical models to study and understand various concepts in fluid dynamics for the past several decades due to several reasons: (i) TC flow is geometrically simple. (ii) It is a closed flow system with exact global balances between the driving and dissipation, and (iii) it is mathematically well defined through the Navier-Stokes equations and the respective boundary conditions. It is thus an ideal playground to study and understand the phenomenon of drag reduction in shear flows under the influence of a dispersed phase like bubbles. In this work we perform two-way coupled numerical simulations of a two-phase TC system to investigate the effects of the dispersed phase on the dynamics of the carrier phase and its consequences on the torque required to drive the system. 

\begin{figure}
  \centerline{\includegraphics[scale=0.45]{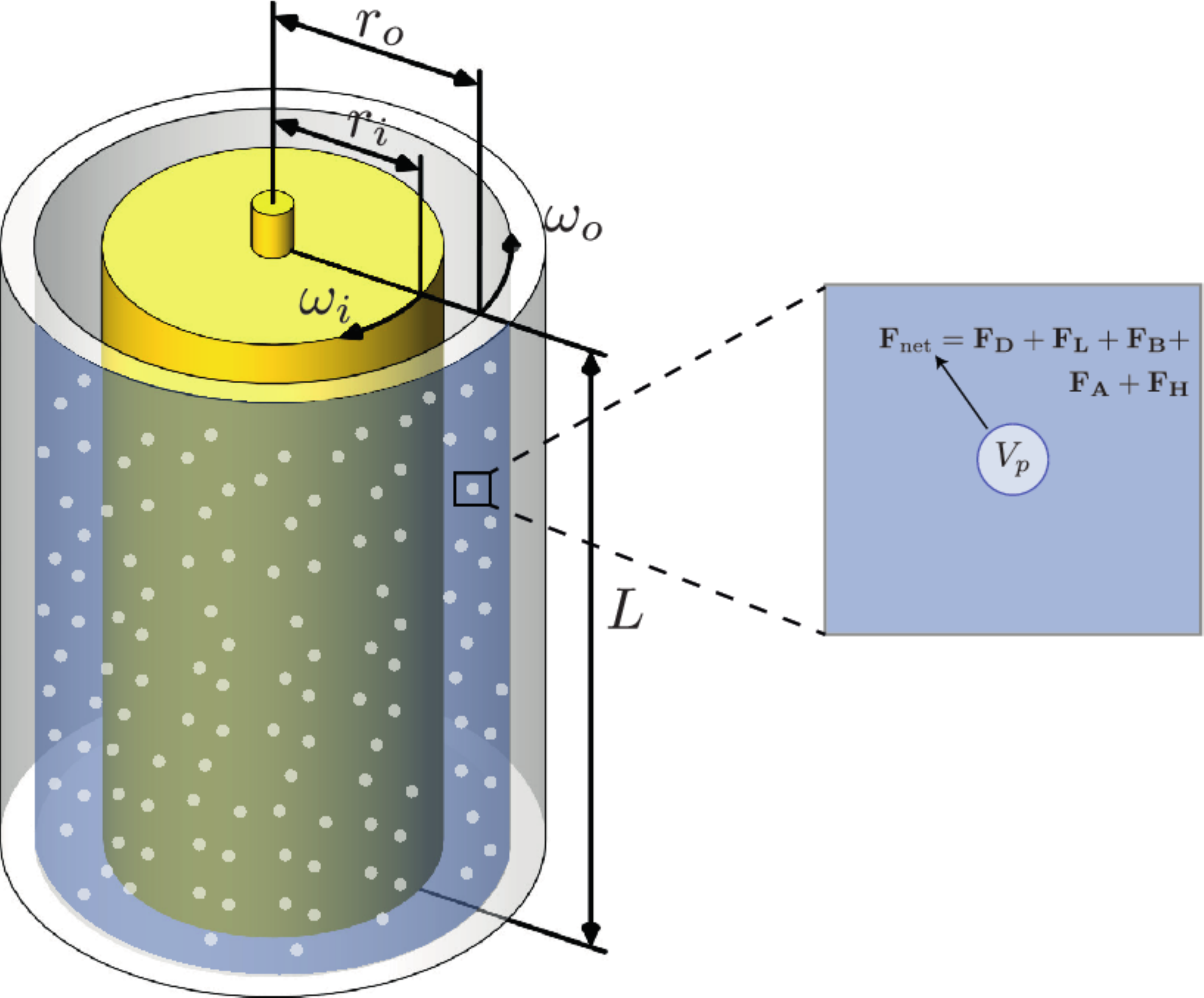}}
  \caption{Schematic of a two-phase Taylor-Couette system with a dispersed phase (bubbles or drops) in the gap width. Setup consists of two co-axial cylinders of length $L$ with an inner cylinder of radius $r_i$ rotating with an angular velocity $\omega_i$ and an outer cylinder of radius $r_o$ rotating with an angular velocity $\omega_o$.}
\label{fig:schem}
\end{figure}

A TC setup consists of two co-axial independently rotating cylinders with a fluid confined between them. The geometry of the TC system can be described using the radius ratio $\eta=r_i/r_o$ and the aspect ratio $\Gamma=L/d$, where $r_i$, $r_o$ are the radius of the inner and outer cylinders respectively, $d=r_o-r_i$ is the gap width and $L$ is the height of the cylinders. Dimensionless radial and axial distances are defined in the form of $\tilde{r}=(r-r_i)/d$ and $\tilde{z}=z/d$, respectively. The driving of the TC system can be expressed in terms of the Reynolds number of the inner and outer cylinders defined by $Re_i=r_i \omega_i d/ \nu$ and $Re_o=r_o\omega_od/\nu$, respectively. In these expressions $\omega_i$, $\omega_o$ are the angular velocities of the inner and outer cylinders, respectively, and $\nu$ is the kinematic viscosity of the carrier fluid confined in the gap width. In this study the outer cylinder is kept stationary and only the inner cylinder is rotated ($Re_o=0$). 

For a single phase TC system, at very low driving the flow is purely azimuthal and in a laminar state. Once the driving of the flow is stronger than a critical value, the purely azimuthal, stable laminar flow is disrupted, leading to the formation of large scale Taylor rolls. Increasing the driving further leads to the onset of interesting flow regimes such as Taylor vortex flow, wavy vortex flow, modulated wavy vortex flow and finally into turbulent Taylor vortices; a detailed phase diagram of these regimes is given by \citet{andereck1986flow}.

The global response of the TC system to the driving can be quantified in terms of the friction factor \begin{equation}
C_f= {\tau_w \over {1\over 2} \rho U_i^2},
\label{eqn:ff}
\end{equation}  
where $\tau_w$ and $U_i$ are the averaged wall shear stress and velocity of the inner cylinder, respectively.  
The friction factor can be seen as the dimensionless drag of the system on the inner cylinder. For pipe
flow the corresponding friction factor
 is the most common way to express the wall drag in dimensionless form \citep{pope2000turbulent}. For Taylor-Couette flow,
alternatively, the response of the system can 
 be quantified in terms of a generalised Nusselt number $Nu_\omega$, which is the angular velocity transport from the inner to the outer cylinder, non-dimensionalized by 
 its value for laminar flow $J_{\mbox{lam}}$ \citep{eckhardt2007torque}, i.e.,
\begin{equation} 
Nu_\omega={J\over J_{\mbox{lam}}}, \qquad\hbox{with} \qquad 
 J=r^3(\langle u_r\omega \rangle_{A,t}-\nu \partial_r \langle\omega\rangle_{A,t}), \label{j-def}
 \end{equation}
 where $\langle ... \rangle_{A,t}$ represents averaging in the two homogenous (azimuthal and axial) directions and also in time and $J_{\mbox{lam}}=2\nu r_i^2r_o^2(\omega_1-\omega_2)/(r_o^2-r_i^2)$.
The relation between the Nusselt number and the friction factor reads

\begin{equation}
C_f = \frac{2\pi L Nu_\omega J_{\mbox{lam}}}{\frac{1}{2}r_iU_i^2}. 
\label{eqn:cf}
\end{equation} 
A recent review on single phase TC flow with the corresponding phase spaces is given by \citet{grossman2016high}. 

In a two-phase TC system a secondary phase is dispersed into the carrier phase (here bubbles dispersed into water) as shown in the schematic of figure \ref{fig:schem}. The dispersed phase is transported by virtue of various forces acing on them such as buoyancy ($F_B$), drag ($F_D$), lift ($F_L$), added mass ($F_A$), history forces ($F_H$) etc. The dispersed phase in turn has a back-reaction force on the carrier flow, thus affecting the angular velocity transport between the two cylinders. This may lead to a change in the torque required to drive the system. A net percentage drag reduction ($DR$) for a two-phase TC system can be quantified according to equation (\ref{eqn_dr}).  

\begin{equation}
\textrm{DR}=\frac{{\langle Nu_\omega\rangle}_s-{\langle Nu_\omega\rangle}_t}{{\langle Nu_\omega\rangle}_s}\times100=\frac{{\langle C_f\rangle}_s-{\langle C_f\rangle}_t}{{\langle C_f\rangle}_s}\times100
\label{eqn_dr}
\end{equation} 
${\langle ...\rangle}_s$ and ${\langle ...\rangle}_t$ correspond to single phase and two-phase systems, respectively. 

The control parameters for the dispersed phase in the case of bubbles are the bubble diameter ($d_{b}$), density ratio (in the case of bubbles dispersed into water $\rho_p/\rho_f\ll 1$), global volume fraction ($\alpha_g=N_pV_p/V$, where $N_p$ is total number of dispersed particles, $V_p$ is the volume of individual particle and $V=\pi({r_o}^2-{r_i}^2)L$ is the total volume of the TC system). In dimensionless form the control parameter can be expressed as Froude number $Fr=\omega_i\sqrt{r_i/g}$ which is the square root of the ratio of centripetal and the gravitational accelerations. Since it is a ratio of forces in two different directions, it does not describe the equilibrium position of the bubbles in the TC gap width. However, it can be interpreted as the relative strength of the buoyancy of bubbles as compared to the driving in the TC system, as has been done in previous studies \citep{murai2005bubble,murai2008frictional,yoshida2009mode,watamura2013intensified}.

\citet{shiomi1993two} performed one of the earliest experiments on two-phase TC flow by studying the various flow patterns that develop in a two-phase mixture confined in a concentric annulus where they observed various flow patterns such as dispersed bubbly, single spiral, double spiral, and triple spiral flows. \citet{djeridi1999bubble} studied the bubble capture and migration patterns in a TC cell for two different configurations, namely (i) with a free upper surface and (ii) with a top stationary wall. In a subsequent study \citet{djeridi2004two} found different spatial structures while using air bubbles and cavitation bubbles separately as the dispersed phase. The focus of these studies was primarily on understanding the different flow patterns that developed in a two-phase TC system, but not on the modification of the torque required to drive the cylinder. \citet{murai2005bubble,murai2008frictional} demonstrated experimentally that a tiny percentage of the dispersed phase (0.1\% volume fraction of bubbles dispersed in silicone oil) can reduce the driving torque on the inner cylinder up to 25\%. The maximum inner cylinder Reynolds number reached in these experiments was $Re_i=4500$ and they found that the overall drag reduction decreased with increasing Reynolds number with almost negligible drag reduction at $Re_i=4000$. With the help of particle tracking velocimetry (PTV) \citet{yoshida2009mode} studied the relationship between the observed drag reduction and changes in the vortical structures of the bubbly TC system. More recently \citet{watamura2013intensified} investigated the effect of micro-bubbles on the properties of azimuthal waves found in TC flow ($\langle Re_i\rangle_{max}$=1000). \citet{fokoua2015effect} performed experiments on a small gap width ($\eta=0.91$) two-phase TC system to study the correlation between bubble arrangement, wavelength of the Taylor vortices, and torque modification. They found that the buoyancy of the bubbles plays a crucial role in the behaviour of the two-phase TC system, thus leading to either drag reduction or drag enhancement. 

In the highly turbulent regime ($Re_i=10^5-10^6$), drag reduction of up to 25\% was achieved by injecting millimetric sized bubbles into the TC flow by \citet{van2005drag,van2007bubbly}. In this regime, it has been demonstrated that deformability of the bubbles can be a deciding factor in achieving a strong drag reduction of up to 40\% (van Gils et al. (2013)). The torque required to drive the inner cylinder in a two-phase TC system (or a bubbly-TC system) depends on various control parameters as mentioned previously. However, in experiments it is not possible to control all these parameters independently, thus making it challenging to study the individual effect of each control parameter on the system. For example as shown in experiments by \citet{murai2005bubble,murai2008frictional}, the Froude number $Fr$, (which is directly related to the velocity of the inner cylinder) cannot be controlled independently from the inner cylinder Reynolds number ($Re_i$). Additionally the diameter of the bubbles dispersed into the gap width depends on the local shear in the flow at the point of injection of the bubbles. In contrast to experiments, numerical simulations allow for more flexible control over the parameters and to explore the underlying physics in detail. 

While there have been a number of experimental studies demonstrating the effect of bubbles on the drag in two-phase TC system, numerical studies have been fairly limited. The computational burden of fully resolving the interactions between thousands of bubbles and the carrier phase is massive, thus making it extremely difficult to simulate thousands of fully resolved bubbles in a highly turbulent carrier flow \citep{lu2005effect,lu2008effect,dabiri2013transition,tryggvason2013multiscale}. In previous studies, this problem has been dealt with by assuming the bubbles to be point-wise sources of momentum and advecting them in a Lagrangian manner \citep{mazzitelli2003effect,sugiyama2008microbubbly,chouippe2014numerical}. Such an approach constrains the size of the particle to the length of the Kolmogorov scale in the carrier flow. Using one-way coupling simulations \citet{climent2007preferential} analysed the migration of point-like bubbles in the Taylor vortex flow and wavy vortex flow regimes. More recently, \citet{chouippe2014numerical} numerically studied the dispersion of such point-like bubbles in a TC system in which the maximum inner cylinder Reynolds number was $Re_i=5000$. However, one-way coupling simulations involve only advecting the dispersed phase based on the local flow conditions in the carrier phase without any back-reaction from the dispersed phase onto the carrier phase. This prevents investigation of any form of drag modification mechanisms that maybe active in a two-phase TC system. This problem is overcome by two-way coupled simulations, namely by implementing the momentum exchange between the bubbles and the carrier phase, while ensuring that the simulations are grid independent and not prone to numerical instabilities. 

Two-way coupling simulations are crucial in order to investigate the effect of a dispersed phase, such as bubbles on the dynamics of a carrier phase. In the context of homogenous isotropic turbulence it has already been shown through two-way coupled simulations, that at large scales the point-like bubbles reduce the intensity in the turbulence spectrum as compared to one-way coupling simulations \citep{mazzitelli2003effect}. Using a two-way coupled Euler-Lagrange scheme, \citet{sugiyama2008microbubbly} reproduced the drag reduction observed in experiments by \citet{murai2005bubble} and concluded that the microbubbles influence the TC system by disrupting the coherent vortical structures and found that the drag reduction decreased at higher $Re_i$; in those simulations the maximum $Re_i$ reached was 2500 where they observed a drag reduction of about 5\%. 

In this present work we use a similar approach by simulating the carrier phase using DNS and tracking the dispersed phase in a Lagrangian manner along with two-way coupling between the phases to extend the $Re_i$ limit to 8000 and gain more insight into two-phase TC drag reduction. We want to find out the effect of bubbles on the driving torque at these higher $Re_i$. The experiments by \citet{murai2005bubble,murai2008frictional} found almost negligible torque modification at $Re_i$=4000, but without independent control over the Froude number $Fr$ due to its implicit dependence on the angular frequency of the inner cylinder and consequently on the Reynolds number $Re_i$. Will the drag reduction be sustained if $Fr$ of the bubbles is controlled or would there be drag enhancement? How does the trajectory of a bubble depend on the level of turbulence in the system and the $Fr$ number? These are the questions we attempt to answer in this work. 
\begin{figure}
  \centerline{\includegraphics[scale=0.45]{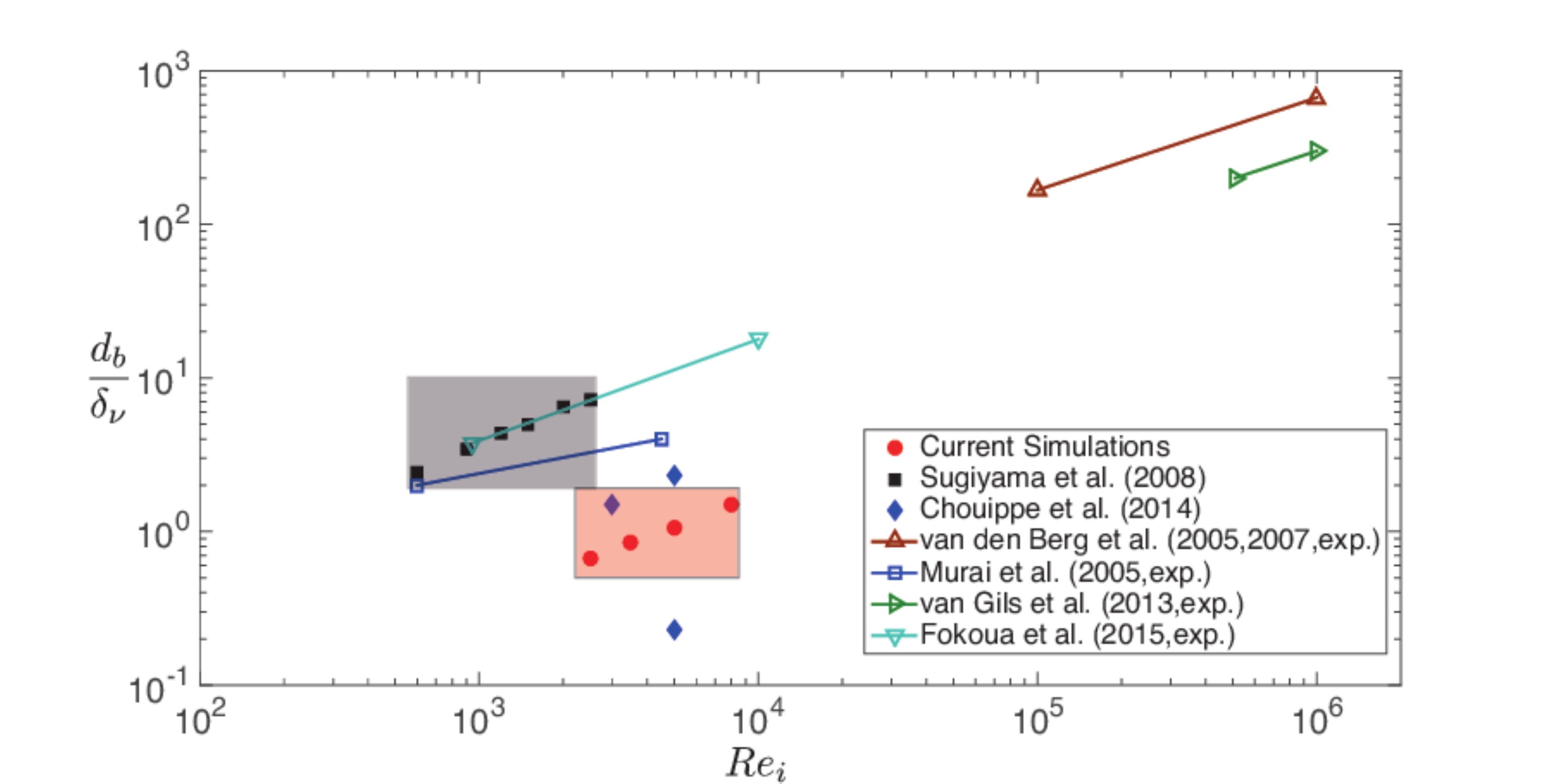}}
  \caption{Phase space of two-phase Taylor Couette; ratio of bubble diameter $d_{b}$ to the viscous length scale $\delta_{\nu}$ is plotted against the inner cylinder Reynolds number $Re_i$. Filled symbols refer to numerical simulations; hollow symbols refer to experimental studies while the lines between the markers indicate a series of experiments. The shaded grey (darker) area indicates the regime of previous two-way coupled simulations on a two-phase TC system, while the shaded orange (lighter) area refers to the current simulations.}
\label{fig:phase}
\end{figure}
Figure \ref{fig:phase} shows the phase space of a two-phase TC system where the ratio of the bubble diameter $d_{b}$ to the viscous length scale $\delta_\nu$ is plotted against the operating Reynolds number $Re_i$ of the inner cylinder. The markers refer to the studies which explore the effect of bubble dispersion and torque modification in the TC system. There are two distinct groupings of studies seen in Figure \ref{fig:phase}; one where the relative particle size $d_b/\delta_{\nu}$ is of the order $O$(1-10) and the rest where $d_{b}/\delta_{\nu}$ is of the order $O(10^2-10^3)$, which correspond to two-phase TC experiments in the highly turbulent regime. The solid markers and lines are numerical simulations and experiments on a two-phase TC system, respectively. The shaded grey area refer to previous two-way coupled simulations of a two-phase TC system \citep{sugiyama2008microbubbly}, while the shaded orange area refers to the $Re_i$ range of current simulations. 

This paper is organised as follows. In Section \ref{sec:eqns}, the governing equations for the carrier phase and the dispersed phase along with the numerical details are given. Section \ref{sec:results} contains the results of this work: We analyse the net drag reduction, contour plots of mean velocity and fluctuations, bubble trajectories and mean distribution of bubbles in the radial and axial directions. In Section \ref{sec:summ} we present a summary of the work along with an outlook for further studies. 
   
\section{Governing Equations and Numerical Details}\label{sec:goveqns}
\label{sec:eqns}
\subsection{Carrier phase}

The dynamics of the carrier phase is computed by solving the Navier-Stokes equations in cylindrical coordinates. For a stationary outer cylinder, i.e. $\omega_o=0$, the governing equations read:

\begin{equation}
\frac{\partial \textbf u}{\partial t}+\textbf u \cdot \nabla \textbf u=-\nabla p+\nu\Delta \textbf u +\textbf f_b(\textbf x,t),
\label{eqn:ns}
\end{equation}
\begin{equation}
\nabla \cdot \textbf u=0 .
\label{eqn:con}
\end{equation} 
In equation (\ref{eqn:ns}), $\textbf f_b(\textbf x,t)$ is the back-reaction force from the dispersed phase onto the carrier phase; this term is ignored in a one-way coupling simulation. In a two-way coupling simulation $\textbf f_b(\textbf x,t)$ is calculated according to \citep{prosperetti2007computational,mazzitelli2003effect,sugiyama2008microbubbly,oresta2009heat}.

\begin{equation}
\textbf f_b(\textbf x,t)=\sum_{i=0}^{N_p} \left(\frac{D\textbf u}{Dt}-\textbf g\right) V_p\delta (\textbf x -\textbf x_p(t)) .
\label{eqn:brf}
\end{equation}
The back-reaction force $\textbf f_b(\textbf x,t)$ is calculated at the exact position of each particle ($\textbf x_p(t)$) which does not necessarily coincide with the location of grid nodes on a fixed Eulerian mesh. A conservative extrapolation scheme is thus necessary to distribute $\textbf f_b(\textbf x,t)$ from the particle position onto the Eulerian grid which will be discussed later. 

For a fully coupled true two-way coupled numerical simulation, we would have to take into account the spatial and temporal evolution of the local bubble concentration $C(x,t)$ in the continuity equation. The continuity equation would then be expressed as $\partial_t(1-C)+\partial_j[(1-C)U_j]=0$ \citep{ferrante2004physical}. However, in our simulations we consider extremely low volume fractions of bubbles (0.1 \%); the maximum instantaneous local volume fraction of bubbles is approximately ten times the global volume fraction 
($\langle C(x,t) \rangle_{\mbox{max}} \sim 1\%$) and in such a case the error in equation \ref{eqn:con}  is negligible. In cases of bubbly turbulent systems with higher global volume fractions, $C(x,t)$ can be much higher than 1\% where the bubbles may interact, coalesce, and can form dense groups of bubbles in the flow.
 In such a case the volumetric effect of the bubbles cannot be neglected anymore.  

DNS of the carrier phase governing equations are performed by numerically integrating the Navier-Stokes equations with a second-order accurate finite-difference scheme \citep{verzicco1996finite, van2015pencil}. The equations are integrated in time using a fractional-step approach. This code has been tested and used extensively in the past to study the dynamics of various turbulent flow states in a single phase TC system for both rotating and stationary outer cylinder configurations \citep{ostilla2013optimal,ostilla2014optimal,ostilla2014boundary,ostilla2014exploring}. Grid nodes are spaced uniformly in the azimuthal and axial directions where the boundary conditions are periodic, while a clipped Chebychev type clustering is employed for spacing the radial nodes. 

\subsection{Dispersed phase}
The bubbles dispersed into the TC system are assumed to be clean, non-deformable and are tracked in a Lagrangian manner with effective forces such as drag, lift, added mass and buoyancy acting on them \citep{mazzitelli2003effect,sugiyama2008microbubbly,oresta2009heat,lakkaraju2014bubbling,chouippe2014numerical}. The generalised momentum equation for particles of density $\rho_p$ dispersed into a carrier fluid of density $\rho_f$ which takes into account the drag, lift, added mass and buoyancy forces reads:

\begin{eqnarray}
\rho_pV_p\frac{d\textbf v}{dt}=(\rho_p-\rho_f)V_p\textbf g-C_D\frac{\pi d_b^2}{8}\rho_f|\textbf v-\textbf u|(\textbf v-\textbf u)+\rho_f V_p C_M \left(\frac{D\textbf u}{Dt}-\frac{d\textbf v}{dt}\right) \nonumber\\
                   + \rho_fV_p\frac{D\textbf u}{Dt}-C_L\rho_fV_p(\textbf v-\textbf u)\times \mathbf{\omega}.
\label{eqn:pmom}
\end{eqnarray}
Based on the particle velocity, the position of each particle is updated according to

\begin{equation}
\frac{d \textbf x_p}{dt}=\textbf v.
\label{eqn:ppos}
\end{equation}
Since we consider clean spherical bubbles the effect of history forces is assumed negligible \citep{magnaudet2000motion,mazzitelli2003effect,sugiyama2008microbubbly}. In equations (\ref{eqn:pmom}) and (\ref{eqn:ppos}), $\textbf u$ is the velocity of the carrier phase at the particle position, $\textbf v$ is the velocity of individual particle, $\textbf x_p$ is the position of the particle, while $C_D, C_M, C_L$ are the drag, lift and added mass coefficients, respectively. Since we consider only very light particles (bubbles in water, i.e. $\rho_p/\rho_f \ll1$), in this case equation (\ref{eqn:pmom}) can be simplified into

\begin{equation}
C_M\frac{d\textbf v}{dt}=-\textbf g-C_D\frac{\pi d_b^2}{8V_p}\rho_f|\textbf v-\textbf u|(\textbf v-\textbf u) + (1+C_M)\frac{D\textbf u}{Dt}-C_L(\textbf v-\textbf u)\times\mathbf{\omega}.
\label{eqn:lpmom}
\end{equation}

In order to close equation (\ref{eqn:lpmom}), along with the coefficients for drag, lift and added mass (i.e. $C_D$, $C_L$, and $C_M$), information on the velocity, total acceleration and vorticity in the carrier phase is required at the exact location of the bubble. Since we assume that the dispersed phase is composed of clean spherical non-deformable bubbles, the value of the added mass coefficient is $C_M=1/2$ \citep{auton1988force,magnaudet2000motion}. In bubbly turbulent flows the exact value of the lift coefficient is not very well known and could be a source of discrepancy in numerical studies. \citet{climent2007preferential,chouippe2014numerical} use a lift coefficient which is dependent on both the bubble Reynolds number and the local shear intensity. By systematically varying the lift coefficient acting on the bubbles from 0 to 0.5 \citet{sugiyama2008microbubbly} found that the value of the lift coefficient plays a crucial role in the mean bubble distribution. We use $C_L=1/2$ \citep{auton1987lift}, a simplified approach which has been used in many previous studies \citep{mazzitelli2003effect,ferrante2004physical,sugiyama2008microbubbly,oresta2009heat,lakkaraju2014bubbling}. Small changes in the magnitude of the lift coefficient can result in different trajectories of the bubbles but in our simulations we find that the qualitative behaviour of the bubbles remains almost the same.

The drag coefficient $C_D$ is computed for each bubble individually and is dependent on the bubble Reynolds number defined as $Re_b=d_b|\textbf v-\textbf u|/\nu$ \citep{mei1992unsteady,magnaudet2000motion}, namely

\begin{equation}
C_D=\frac{16}{Re_b}\Bigg[1+\frac{Re_b}{8+\frac{1}{2}(Re_b+3.315\sqrt{Re_b})}\Bigg] .
\end{equation}

The difference in the location of the bubbles and the grid nodes of the Eulerian mesh restricts us from calculating local flow quantities (i.e. velocity, total acceleration, and vorticity) directly from the carrier phase solution. Since the velocities are in three different positions in a staggered grid arrangement, all the three velocities belonging to any specific computational cell is first interpolated to the cell centre using the values from the nearest grid points. A tri-linear interpolation scheme, containing information from all the cell-centres surrounding a specific particle is used to calculate the carrier phase velocity at the exact particle position. The same approach is employed for the interpolation of the total acceleration and the vorticity in the flow. The error in the interpolation decreases asymptotically as $(\Delta \textrm x)^2$ as the grid spacing $\Delta \textrm x$ tends to zero \citep{yeung1988algorithm}. 

In addition to interpolating local flow quantities, the back-reaction force from the bubbles (equation \ref{eqn:brf}) needs to be extrapolated to the surrounding grid nodes. The back-reaction force from the particle $\textbf f_b(\textbf x,t)$ is extrapolated to the surrounding grid nodes residing in fixed computational volume (which is smaller than the bubble volume) using an exponential distribution function, which decreases monotonically with the distance between the grid nodes and the particle position. In the context of simulating particle-laden flows such an approach has already been discussed in detail by \citet{capecelatro2013euler}. The extrapolation scheme ensures second-order accuracy, conservation of the point back-reaction force while extrapolating the data from a Lagrangian location to a Eulerian mesh and is also stable when a non-uniform grid distribution is used. Time integration of equations (\ref{eqn:ppos}) and (\ref{eqn:lpmom}) is done by a second-order accurate Runge-Kutta scheme. The code has been parallelised using MPI and OpenMP directives and has a strong scaling of up to 1000 cores. 

\begin{figure}
  \centerline{\includegraphics[scale=0.55]{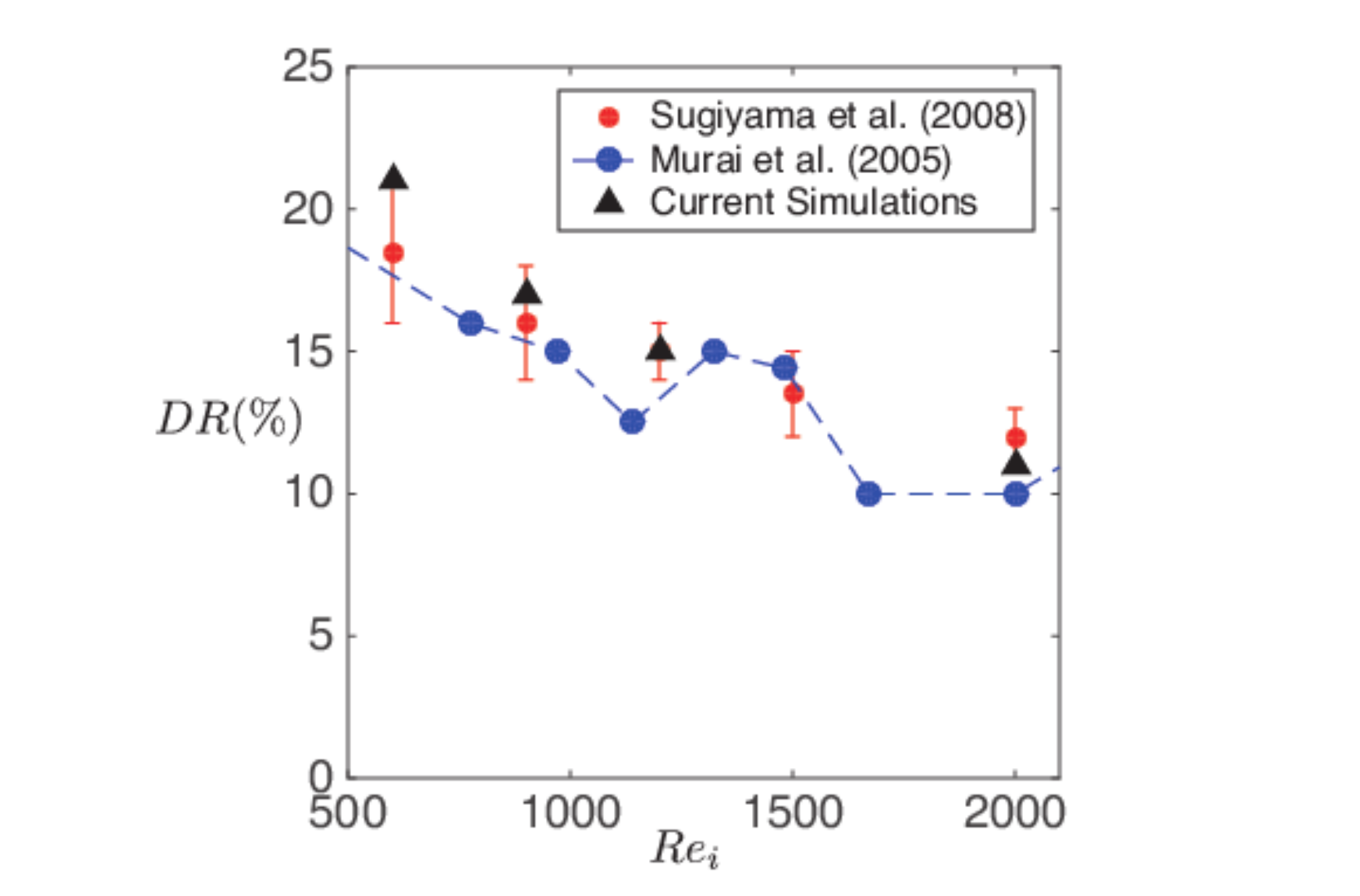}}
  \caption{Percentage of drag reduction ($DR$) as a function of 
   the inner Reynolds number $Re_i$. 
    The results from the current code are compared against the numerical simulations of \citet{sugiyama2008microbubbly} and experimental measurements of \citet{murai2005bubble}. The global volume fraction of bubbles $\alpha_g=0.125\%-0.670\%$ ; $Fr=0.3-1.0$.}
\label{fig:valid}
\end{figure}

In order to validate the two-phase code, we perform simulations with the exact resolutions and parameters as done in \citet{sugiyama2008microbubbly} and compare our results in Figure \ref{fig:valid}. Very good agreement is seen between the results of the current simulations and the simulations of \citet{sugiyama2008microbubbly} and also the experiments of \citet{murai2005bubble} where the drag reduction (DR) is plotted against the inner cylinder Reynolds number. For the rest of the cases simulated in this paper the parameters chosen are as follows: The geometry of the TC system is fixed by fixing the radius ratio $\eta$=0.833 and the aspect ratio to $\Gamma$=4. The global volume fraction $\alpha_g$ is fixed to 0.1 \%, the relative bubble diameter to $d_b/d$=0.01, while the Froude number $Fr$ of the bubbles is varied from 0.16 to 2.56. 

To reduce the computational cost, the simulation is initially run without any dispersed phase and once the flow in the single phase reaches a statistically stationary state, bubbles are placed at random positions throughout the domain. The velocity of these bubbles is equated to that of the velocity of the carrier phase interpolated at the bubble location and the two-phase flow is allowed to develop. Once the bubbles are introduced into the flow, the previously developed statistically stationary state of the single phase flow is disrupted and transients develop in the two-phase flow which eventually die out after approximately 100 full cylinder rotations. Additionally, it has been ensured that the mean axial velocity of the carrier phase in the domain is equal to zero.
Similar to the single phase system, azimuthal and axial periodicity is employed for the dispersed phase i.e. when the bubbles exit an azimuthal or axial boundary they are re-entered at the opposite boundary at the exact same location with the same velocity. This approach is obviously different from experimental studies \citep{murai2005bubble,murai2008frictional,van2013importance,fokoua2015effect}, where the bubbles are injected at the bottom and are collected at the top and can be a cause of discrepancy in the comparison between numerical and experimental data. 

In the current simulations, through some additional test runs we have ensured that the chosen axial extent of the domain does not influence the results. (Details on the Nusselt number for both single phase and two-phase case with $Fr=0.16$ is compared for $\Gamma=4$ and $\Gamma=8$ in table \ref{tab:tordets} of Appendix \ref{app:A}). However, it can be expected that after the two-phase system achieves a statistically stationary state, both the experimental and the computational systems behave in a qualitatively similar manner. An elastic bounce model is used for the interaction of the bubbles with the inner and outer cylinder wall as implemented in previous studies \citep{climent2007preferential,sugiyama2008microbubbly,chouippe2014numerical}. While modifying the elastic bounce interaction may result in a change in the local bubble concentration near the wall, small variations in the coefficient of restitution of bubble impact with the wall does not influence the qualitative behaviour of the motion of bubbles which we study in a later section. Additionally, since the bubbles are assumed to be Lagrangian points, there is a possibility of bubbles overlapping among each other. Owing to the very low global volume fraction in the system ($\alpha_g=0.1\%$), the fraction of bubbles overlapping onto each other when taking into consideration their physical size is less than $10^{-3}$, which we consider to be negligible. 

\begin{table}
  \begin{center}
\def~{\hphantom{0}}
	\begin{tabular}{c|c|c|c|c}
	$Re_i$ & $N_r\times N_\theta \times N_z$ & $\Delta r^+$ & $d_b/\delta_\nu$ & $St=\tau_b/\tau_\eta$ \\
	\hline
	\hline
	2500 & 180$\times$100$\times$150 & 0.125 & 0.55 & 0.02 \\
	3500 & 200$\times$100$\times$180 & 0.150 & 0.75 & 0.04 \\  
    5000 & 220$\times$120$\times$200 & 0.150 & 1.0 & 0.07 \\
    8000 & 258$\times$150$\times$220 & 0.125 & 1.25 & 0.10 \\
    \hline
    \end{tabular}
    \caption{Numerical details of the simulations. First column is the operating Reynolds number of the inner cylinder, second column is the radial grid spacing near the inner wall, the third column shows the radial grid refinement near the inner cylinder, the fourth column shows the diameter of the bubble normalised by the viscous length scale and the fifth column is the bubble Stokes number for each case. These values are for a geometry with radius ratio $\eta=0.833$ with an aspect ratio $\Gamma=4$ and a rotational symmetry $n_{sym}=6$ in the azimuthal direction while the maximum value of the bubble size relative the radial grid refinement in all simulations is $d_b/\Delta r_\textrm{min} \sim 3$}
    \label{tab:radst}
  \end{center}
\end{table} 

\begin{figure}
\centerline{\includegraphics[scale=0.5]{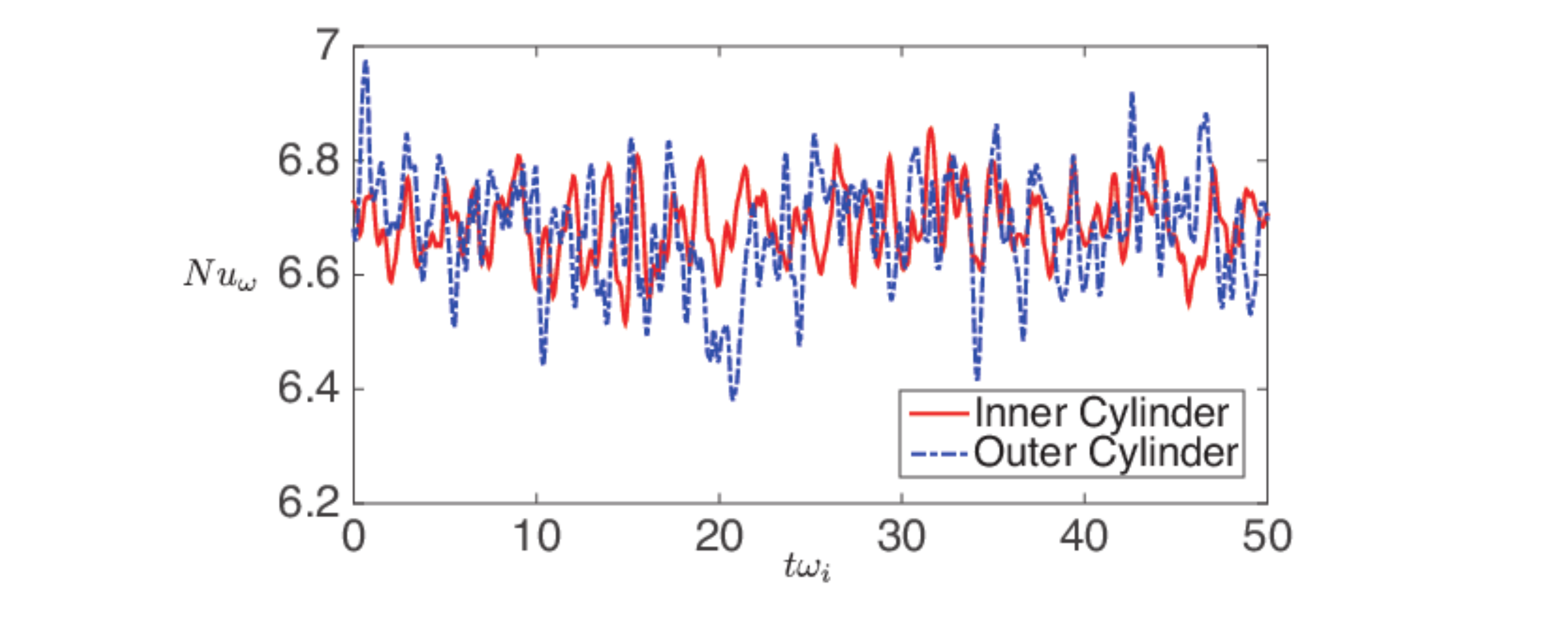}}
\caption{Typical time series of the instantaneous torques at the inner (solid red line) and outer cylinder (dashed blue line). $Re_i$=2500, $Fr$=0.16, $\alpha_g$=0.1 \%  Temporal origin is arbitrary and after the system has achieved statistically stationary state.}
\label{fig:ttime}
\end{figure}

The procedure for calculating the Nusselt number $Nu_\omega$ for both the single phase and the two-phase systems is as follows. The simulation is run for at least 50 full cylinder rotations after statistical stationarity is achieved and convergence is ensured by comparing the mean azimuthal velocity profiles for 50 and 100 full cylinder rotations. Additionally it is ensured that the net angular momentum current in the radial direction is constant. The torques on both the cylinders are then averaged in time and it is ensured that the difference in $\langle Nu_\omega \rangle$ for both inner and outer cylinders is less than 1 \%. Figure \ref{fig:ttime} shows a typical time series of the non-dimensional angular velocity transport $Nu_\omega$ on both the inner and outer cylinder after the transients in the two-phase system die out and the system reaches a statistically stationary state. The net percentage drag reduction ($DR$) is then calculated according to equation (\ref{eqn_dr}). To maintain a global volume fraction of $\alpha_g$=0.1\% with bubbles of size $d_b/d$=0.01, approximately 50000 bubbles are required in the simulation. In order to calculate the effective forces acting on the point-wise particles according to equation (\ref{eqn:lpmom}), and advect them with sufficient accuracy, the size of the bubble should be of the same order as the smallest relevant length scale in the carrier phase. It can be seen from figure \ref{fig:phase} that the ratio of bubble diameter ($d_b$) to the viscous length scale $\delta_\nu$ is of the order of 1 for all the simulations considered in this work. In table \ref{tab:radst} we give details on the resolutions for various Reynolds number and the corresponding Stokes number of the bubbles ($St=\tau_b/\tau_\eta$, where $\tau_b=d_b^2/24\nu$ is the bubble response time and $\tau_\eta$ is the Kolmogorov time scale). In the next Section we discuss the main results of this paper, where we study the effect of the operating Reynolds number $Re_i$, and the Froude number $Fr$ on the global response of the two-phase TC system and also on the local flow dynamics. 

\section{Results}
\label{sec:results}
\subsection{Drag reduction}

\begin{figure}
\centerline{\includegraphics[scale=0.40]{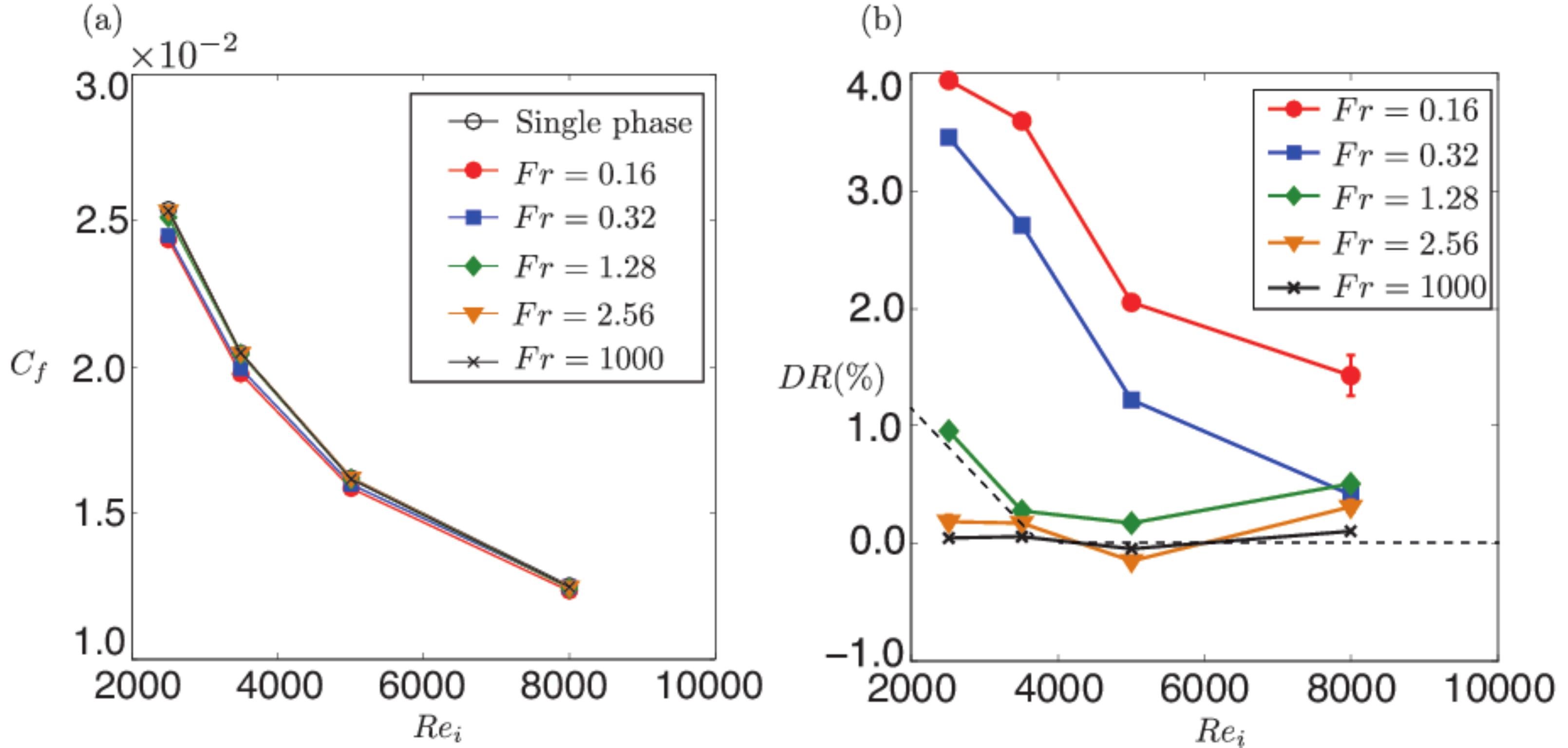}}
\caption{
(a) Friction factor $C_f(Re_i)$ for single phase TC flow and for bubbly TC flow 
  with bubbles of five different Froude numbers.  Only for $Fr<1$ drag reduction occurs. This drag reduction
  is much better expressed  in terms of 
   the percentage of drag reduction ($DR$), which is plotted in (b), again as a function of 
   the inner Reynolds number $Re_i$.
Positive $DR$ indicates that the driving torque in the two-phase case is lower than that of the single phase case. The dashed line shows the path taken by $Re_i$-$Fr$ number in the experiments by \citet{murai2008frictional}.}
\label{fig:dr_re}
\end{figure}

We first focus on the change in the global response of the TC system with the addition of the
dispersed phase. In figure \ref{fig:dr_re}(a) we compare the friction coefficient $C_f$ for the single phase and two-phase cases for five different Froude numbers $Fr$=0.16, 0.32, 1.28, 2.56 and 1000. When $Fr<1$ and 
in particular for low  $Re_i$, we observe a reduction in the friction factor $C_f$ for the 
two-phase as compared to the single flow case or as compared to cases with $Fr \gg 1$. In order to observe this difference more clearly we compute the net percentage drag reduction $DR$ according to equation \ref{eqn_dr}.
It is shown as function of $Re_i$ for the five different $Fr$ in figure \ref{fig:dr_re}(b). Almost 4\% drag reduction is achieved at $Re_i$=2500 and $Fr$=0.16. For fixed and small enough $Re_i$, increasing the $Fr$ number of the bubbles leads to a decrease in the drag reduction. Vice versa when the $Fr$ number is kept fixed the drag reduction $DR$ gets less for increasing $Re_i$. This holds in particular for configurations with $Fr<1$, i.e., when the buoyancy of the bubbles is dominant over the driving of the system. For $Fr>1$, for at least within the examined $Re_i$ range, there is overall no systematic trend with $Re_i$ and negligible drag reduction, thus indicating that bubble buoyancy has a strong role in achieving drag reduction. A unique difference in these results when compared to the experimental study of \citet{murai2008frictional} is that here we have independent control over the $Fr$ number, i.e. independently of the Reynolds number $Re_i$. \citet{murai2008frictional} found almost negligible drag reduction beyond $Re_i$=3000; however in their setup the $Fr$ of the bubbles is dependent on $Re_i$ and at $Re_i=2000$ the $Fr$ number was already above one. The path taken by the $Re_i$-$Fr$ number in the experiments is shown in figure \ref{fig:dr_re}(b) by a dashed line. In the current simulations, in contrast, we observe that by fixing the $Fr$ number of the bubbles to less than one, drag reduction can be observed even at $Re_i$=8000. The details on the Nusselt number for the single phase and two-phase cases for these simulations in given in table \ref{tab:tordets} in Appendix \ref{app:A}.

It can be expected that at the nearly asymptotically large value of $Fr=1000$, the centripetal acceleration can play the role of buoyancy for bubbles immersed in a horizontal channel flow \citep{xu2002numerical,murai2007skin,harleman2012effect,pang2014numerical} which may lead to drag reduction. However, we do not find any such drag reduction in the TC system as seen in figure \ref{fig:dr_re} and we think this might be due to different physical mechanisms governing drag reduction in both systems. In comparison with the current simulations, experiments with larger bubbles and thus higher gas volume fractions ($d_b^+>>1$, $\alpha_g>1\%$) have been performed in (bubbly) turbulent channel flows to obtain sustainable levels of drag reduction \citep{xu2002numerical,murai2007skin}. In experiments with smaller bubbles and lower gas volume fractions ($d_b^+\sim1$, $\alpha_g\sim1\%$), \citet{pang2014numerical} and \citet{harleman2012effect} found extremely small levels of drag reduction which is similar to what we find in our simulations. Additionally, the drag reduction effect observed in Taylor-Couette flows in the low-Reynolds number regime is through disruption of coherent Taylor rolls which are fixed in space and time for single phase flows and are responsible for the majority of the angular momentum transport. In contrast, for channel flows co-spectra analysis shows that these wall-attached large scale structures are inactive \citep{jimenez2011cascades}.

\subsection{Carrier phase velocity fields}

\begin{figure}
\centerline{\includegraphics[scale=0.45]{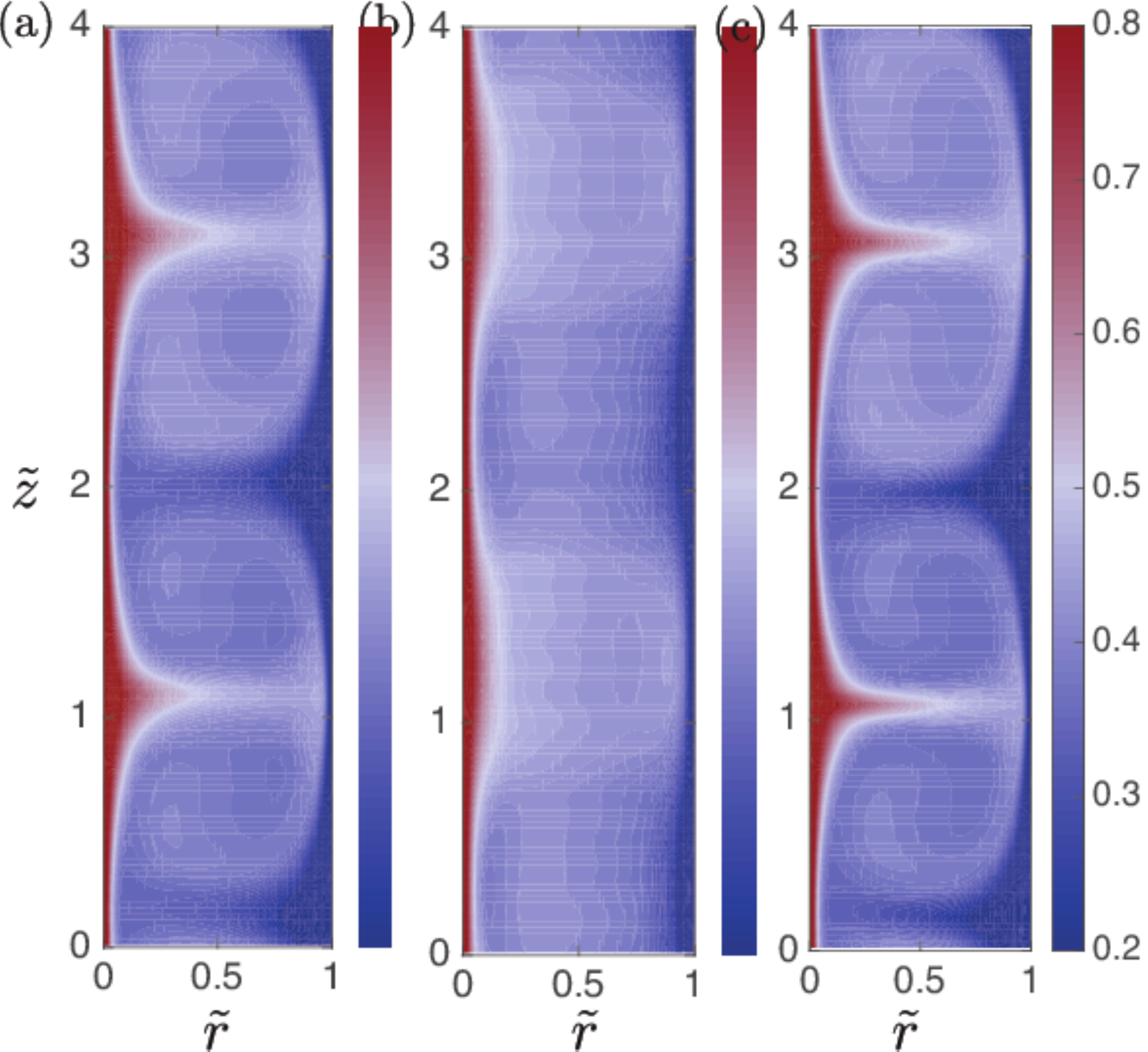}}
\caption{(Colour online) Contour plots of azimuthally- and time averaged azimuthal velocity field $\langle\bar{u}_\theta\rangle_{\theta,t}$ for a fixed $Re_i=2500$ (a) Single phase (b) Two-phase flow with $Fr$=0.16 and (c) with $Fr$=1.28. For the low $Fr$ case the Taylor rolls are considerably weakened by the strongly buoyant bubbles.}
\label{fig:azicon}
\end{figure}

In figure \ref{fig:azicon} we compare the contour plots of the azimuthal velocity, averaged in the azimuthal direction and over time. The Reynolds number is fixed at $Re_i$=2500 and averaged contours of two systems with $Fr$=0.16 and $Fr$=1.28 is compared with that of the single phase flow. As mentioned earlier (also see again figure \ref{fig:dr_re}), the $DR$ decreases with increasing $Fr$ for fixed $Re_i$. In figure \ref{fig:azicon}(c) ($Fr=1.28$), a clear signature of the Taylor vortex can be observed with a structure very similar to that of the single phase case as seen in figure \ref{fig:azicon}(a). When $Fr<1$ (figure \ref{fig:azicon}b) a strong footprint of the Taylor vortex is not observed anymore. The strong buoyancy of the bubbles as compared to the driving of the system (i.e. $Fr<1$) is responsible for disrupting the Taylor vortices, which have concentrated regions of high strain-rates and are thus highly dissipative \citep{sugiyama2008microbubbly}. With increasing $Fr$ the Taylor vortex structure resembles more to that of a single phase system \citep{ostilla2013optimal}, resulting in a drop in $DR$. 

\begin{figure}
\centerline{\includegraphics[scale=0.45]{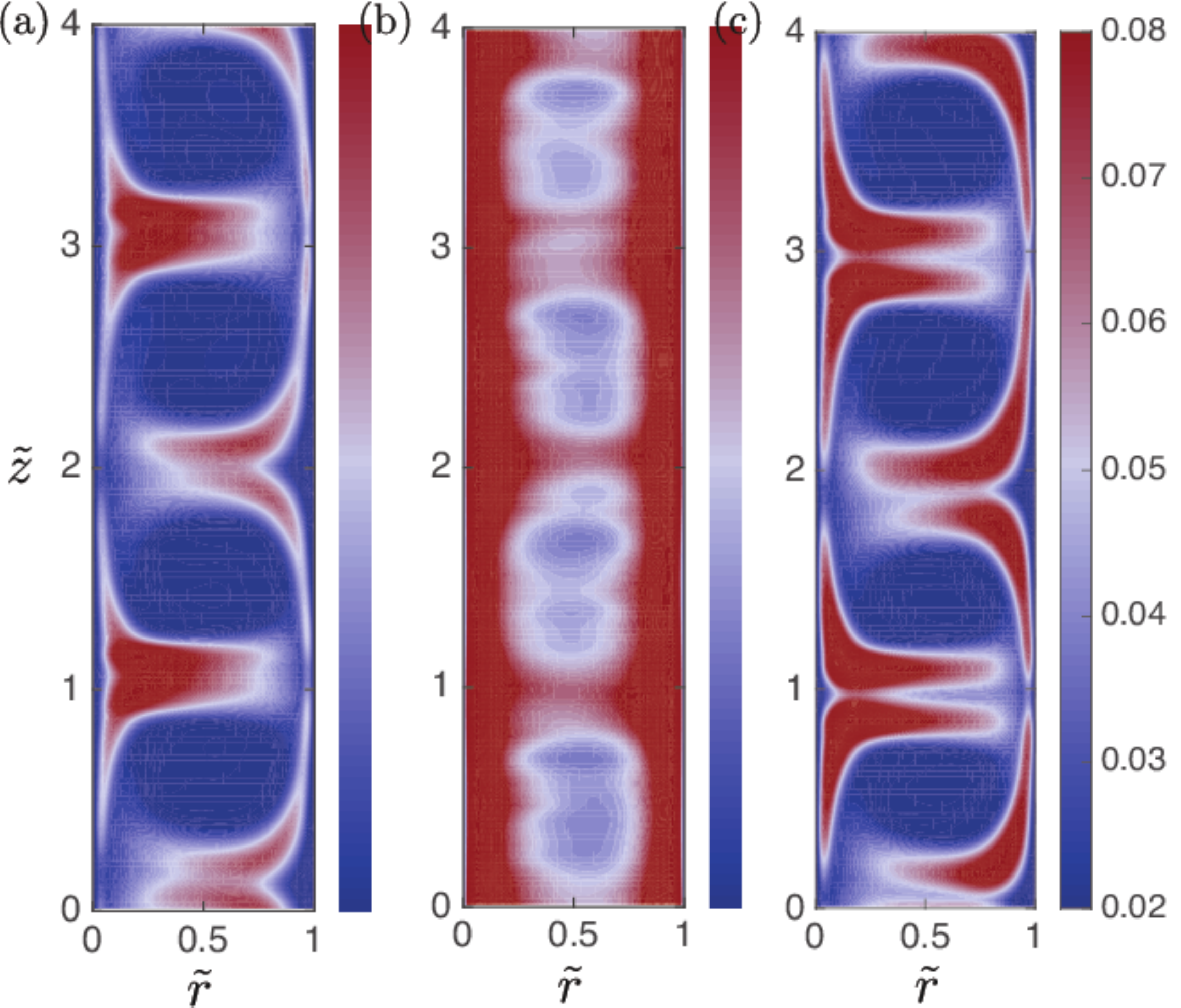}}
\caption{(Colour online) Contour plots of azimuthal and time averaged r.m.s. of the velocity fluctuations ($u'_\theta$) for a fixed at $Re_i=2500$; (a) Single phase (b) Two-phase flow with $Fr$=0.16 and (c) with $Fr$=1.28. Same colour bar is used for all three plots. Again, the effect of the strongly buoyant bubbles ($Fr=0.16$) on the Taylor rolls can be clearly observed.}
\label{fig:azifluc}
\end{figure}

This is also observed in figure \ref{fig:azifluc} where we show contour plots of azimuthally and time averaged r.m.s. of the azimuthal velocity fluctuation $u'_\theta$. These are computed in form of the root mean squared (r.m.s.) value as $u'(r,z)=[\langle u^2\rangle_{\theta,t}-\langle u\rangle^2_{\theta,t}]^{\frac{1}{2}}$. For the single phase flow (figure \ref{fig:azifluc}(a)), there exist two localised regions  near the inner-cylinder which show a peak in the velocity fluctuations. These regions are associated to the sites of plume ejection \citep{ostilla2013optimal,ostilla2014optimal,ostilla2014boundary}. In a two-phase system with strong buoyant bubbles ($Fr<1$ i.e. figure \ref{fig:azifluc}(b)) there is a peak in the fluctuation along the complete axial extent of both the inner and outer cylinders. The bubbles rising near the walls of the cylinders by virtue of their buoyancy are responsible for such a behaviour. However, unlike in the single phase case where the fluctuations extend into the gap width in the two-phase case with $Fr<1$ they are localised near the cylinder walls giving and indication that the plumes are much weaker. This is not observed in the case where the bubbles are weakly buoyant ($Fr>1$, i.e. figure \ref{fig:azifluc}(c)). Instead, we see that the axial extent of the plume ejection site is extended when compared to the single phase system. This is a result of bubbles being captured by the Taylor vortices at the sites of plume ejection after sliding on the inner cylinder for a short period. For a single phase flow, the plumes ejected on the inner cylinder wall are responsible for majority of the angular velocity transport. Additionally, they assist in the transition to the ultimate state of turbulence where the turbulence is fully developed in both bulk and boundary layers \citep{ostilla2014boundary}. 

\begin{figure}
\centerline{\includegraphics[scale=0.5]{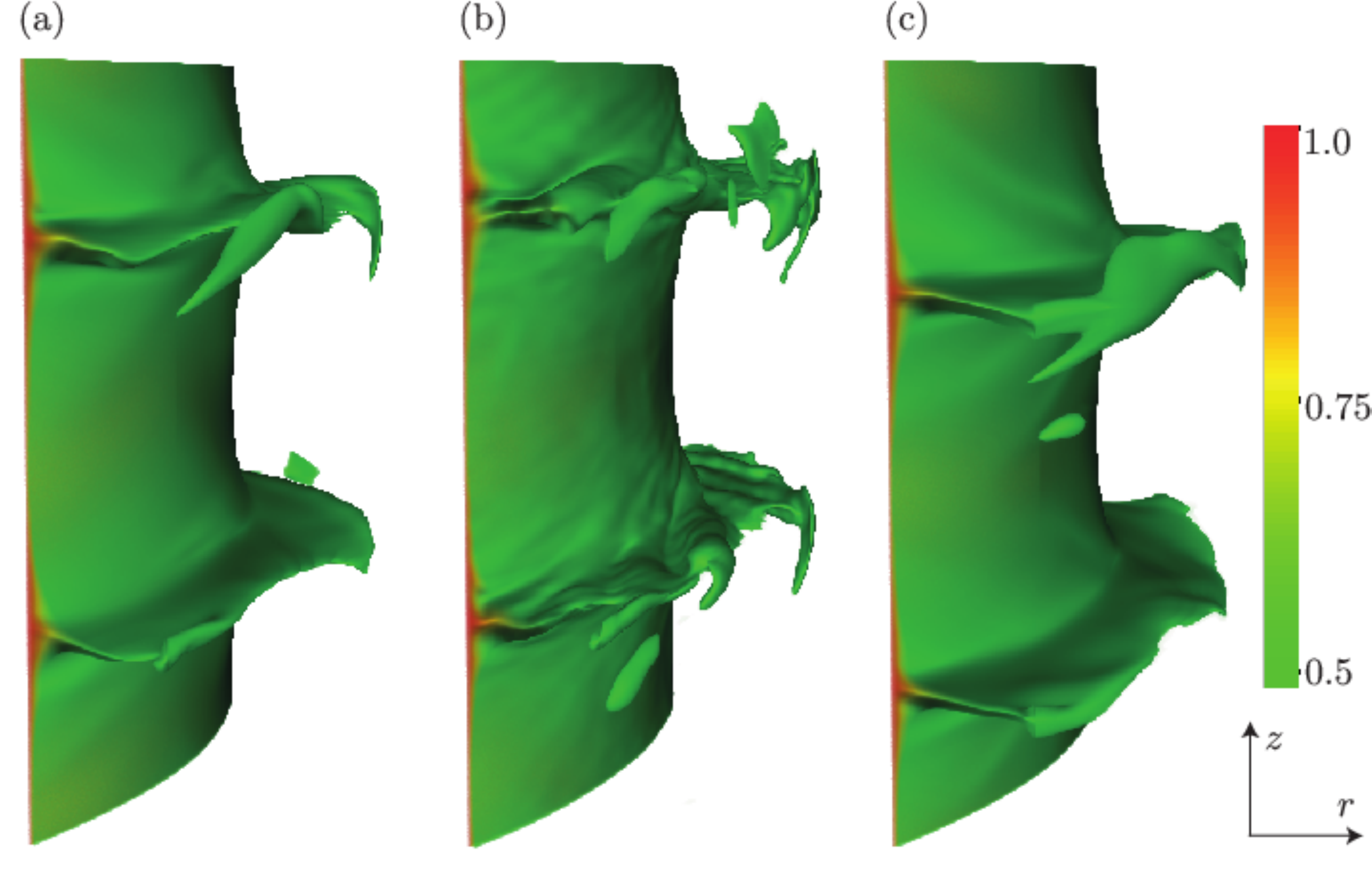}}
\caption{Three dimensional instantaneous snapshot of the azimuthal velocity field $u_\theta$ for a fixed $Re_i=2500$ (a) Single phase (b) $Fr$=0.16 (c) $Fr$=1.28.}
\label{fig:instan}
\end{figure}

\begin{figure}
\centerline{\includegraphics[scale=0.45]{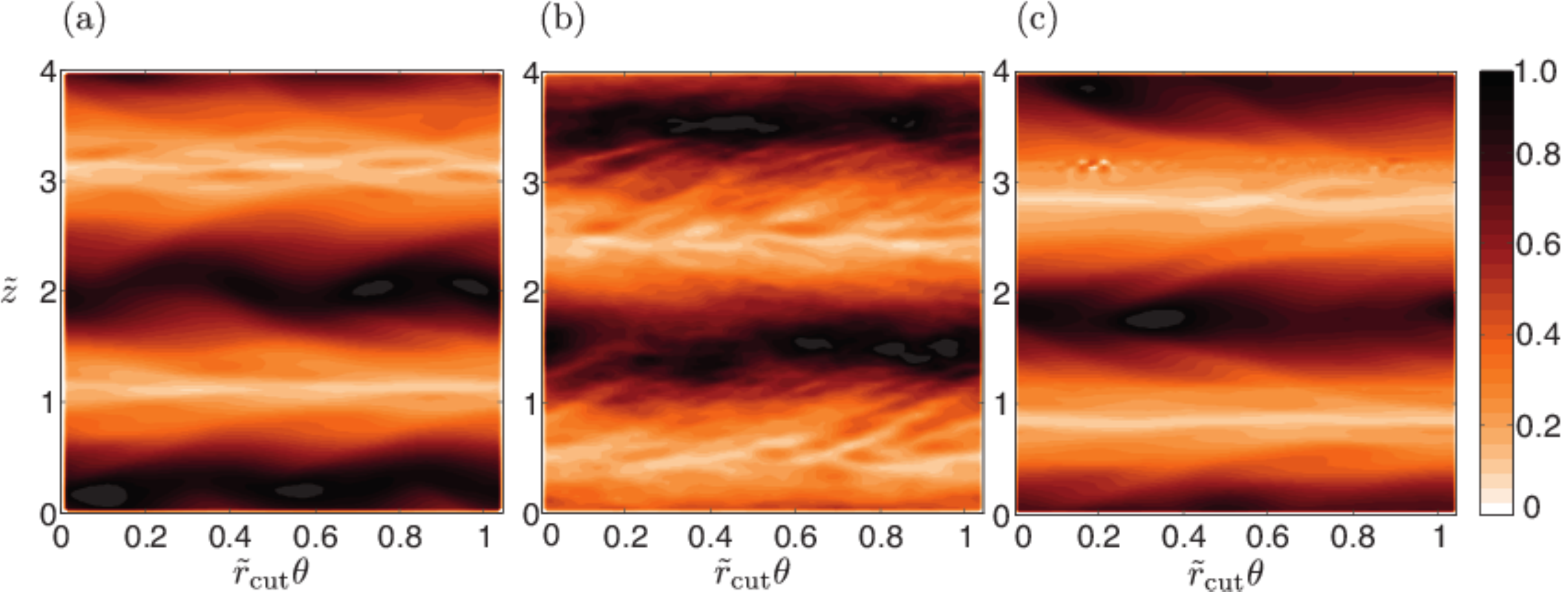}}
\caption{Near wall strain rate contours ($\hat e_{r,\theta}$) at a constant radius cut ($r^+=0.25$) for $Re_i=2500$. Left panel corresponds to the single phase, middle panel to the two-phase case with $Fr=0.16$ while right most panel is with $Fr=1.28$.}
\label{fig:wallstr}
\end{figure}

In figure \ref{fig:instan} we show three dimensional instantaneous snapshots of the azimuthal velocity field for the single phase case and the two-phase cases for both $Fr$=0.16 and $Fr$=1.28. The important observation made here is that while the ejection of plumes from the inner cylinder ($\tilde{r}$=0) is very strong for the single phase case and $Fr=1.28$, it is much weaker for $Fr=0.16$. The strong buoyancy of the bubbles ($Fr=0.16$) causes the plume ejection region to sweep the inner cylinder surface and thus when averaged in time a clear footprint of the ejection region is not observed anymore (c.f. figure \ref{fig:azicon}b). The weakening of the plumes and the Taylor rolls is also clearly observed in figure \ref{fig:wallstr} where we plot contours of the normalised wall-parallel strain rate close to the wall. The local strain rate is normalised using the maximum strain rate for each case and these contour plots are a direct indication of the spatial distribution of the wall shear stress on the inner cylinder. As seen in \citet{ostilla2014boundary}, the local inner cylinder boundary layer profiles vary depending on whether the region belongs to a plume ejection site or a plume impacting site. This is also reflected in the contours shown in figure \ref{fig:wallstr}, where in the case of single phase flow high $e_{r,\theta}$ regions can be observed at the plume impacting sites. The strain rates in these sites are comparatively weaker in the case of $Fr=0.16$ due to the loss of strength in the plume ejections and as a consequence also the coherent Taylor rolls. With increase in $Fr$ numbers which is shown in the right most panel of figure \ref{fig:wallstr}, the plumes ejections regain their strength and the flow becomes similar to the single phase case with bubbles having a much weaker effect.

\begin{figure}
\centerline{\includegraphics[scale=0.6]{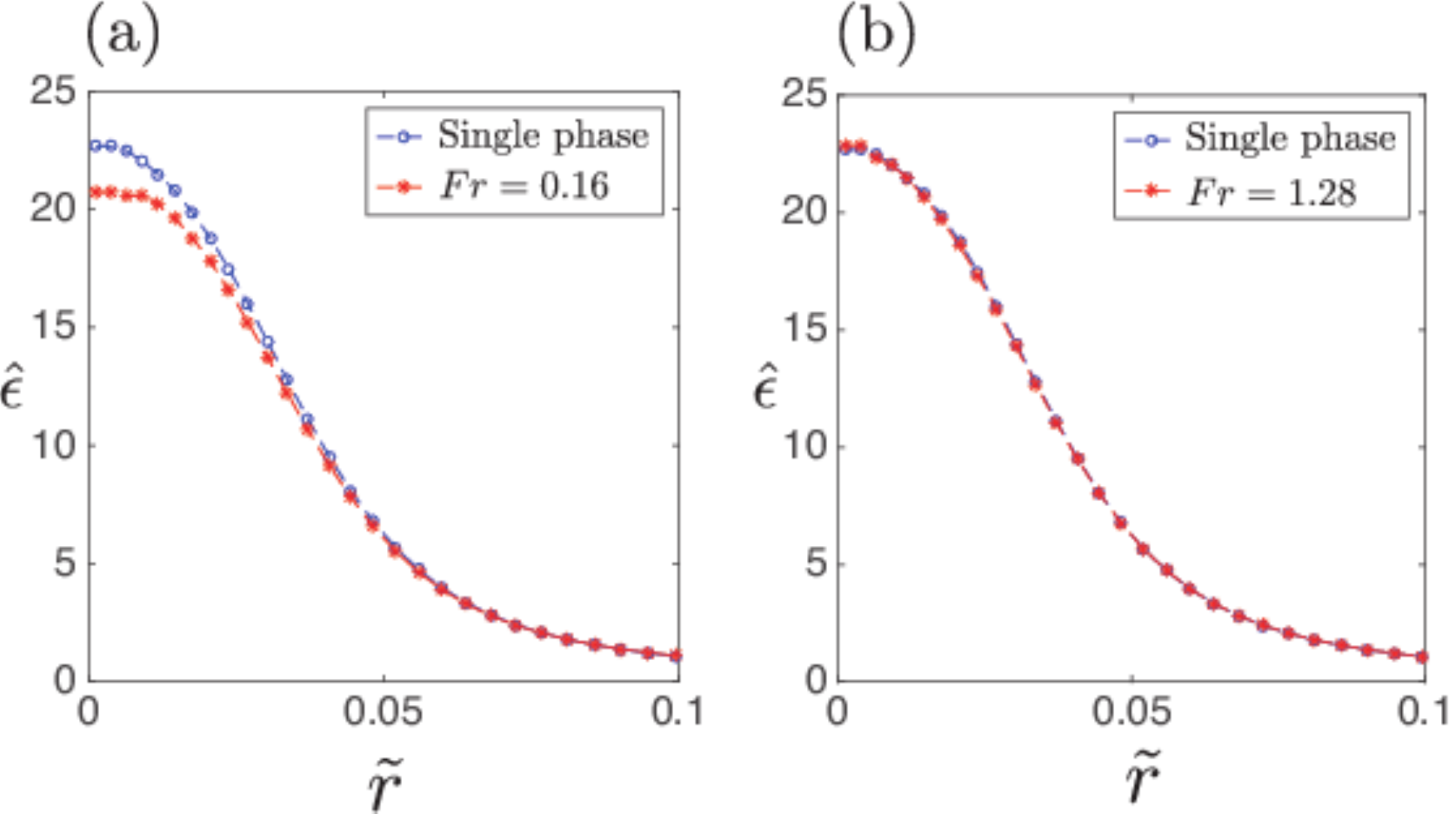}}
\caption{Comparison of azimuthal, axial and time averaged viscous dissipation profiles near the inner cylinder wall. The $Re_i$ in both (a) and (b) is 2500. For the two-phase system (a) $Fr$=0.16 (b) $Fr$=1.28}
\label{fig:radis2500}
\end{figure}

In order to understand more on how the bubbles affect these plumes, in figure \ref{fig:radis2500} we plot the viscous dissipation profile near the inner cylinder wall. The total dissipation per unit mass is calculated as a volume and time average of the local dissipation rate, and as shown by \citet{eckhardt2007torque} is directly related to the driving torque. The viscous dissipation is normalised accordingly $\hat{epsilon}=\epsilon/\nu(U_i/d)^2$ and in figure \ref{fig:radis2500}, we compare the dissipation profiles of a two-phase system with strongly buoyant bubbles ($Fr<1$) and weakly buoyant bubbles ($Fr>1$) with a single phase system. When $Fr<1$, the viscous dissipation is lower than that compared to the single phase and difference can be clearly observed in figure \ref{fig:radis2500}(a). This is not the case in figure \ref{fig:radis2500}(b), where the difference is almost negligible and thus also resulting in minimal drag reduction. Such a behaviour is related to the pattern of distribution of bubbles near the inner cylinder and is discussed in a later section. In addition to a drop in $DR$ with increasing $Fr$ at a fixed $Re_i$, we also notice a drop in $DR$ with $Re_i$ even for bubbles with $Fr<1$. With increase in $Re_i$, the coherent flow structures become less responsible for the angular velocity transport \citep{ostilla2013optimal,ostilla2014boundary}. The strong buoyancy effect of the bubbles (i.e. $Fr<1$) thus acts on weaker coherent structures and lose their efficiency on affecting the angular velocity transport in the system. This trend was also observed in the experiments by \citet{murai2005bubble,murai2008frictional} who found that the $DR$ dropped to almost zero at $Re_i$=4000. We show that although there is a drop in $DR$ with increasing $Re_i$, little $DR$ can be attained even at $Re_i$=8000 provided $Fr<1$. 

\subsection{Motion of dispersed phase}
\label{sec:motdis}
\begin{figure}
\centerline{\includegraphics[scale=0.5]{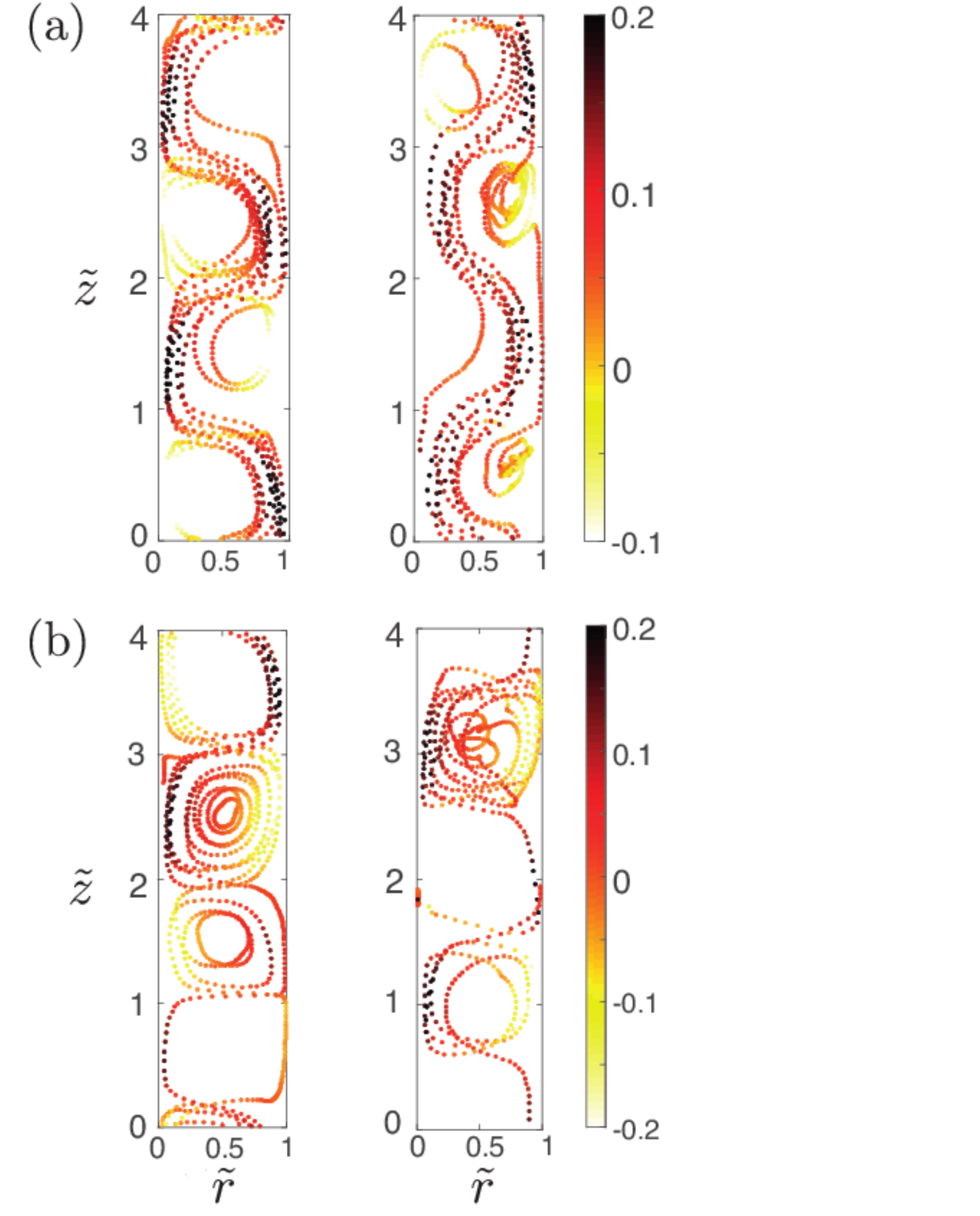}}
\caption{Full trajectory of a single bubble in the $r$-$z$ plane of the two-phase TC system.  Left panels in both (a) and (b) belong to $Re_i$=2500 while the right panels are for $Re_i$=5000. $Fr$ number for the bubbles in the top panels i.e. (a) $Fr$=0.16, bottom panels (b) $Fr$=1.28. The bubble is tracked for almost 20 full inner cylinder rotations in all cases after the system has reached statistically stationary state. Same colour-scale is used for cases with the same $Fr$ numbers and indicates the axial velocity $v_z$ of the bubbles. While the less buoyant bubbles ($Fr=1.28$, (b)) tend to get trapped in the Taylor rolls, the more buoyant ones ($Fr=0.16$, (a)) rise through and weaken them.}
\label{fig:parttraj}
\end{figure}

In this section we look at the different trajectories of bubbles for different $Re_i$ and $Fr$ numbers. In figure \ref{fig:parttraj}, we show four different trajectories of an individual bubble for two different $Re_i$ and $Fr$ numbers, respectively. There is a strong azimuthal motion imposed on the bubbles by the carrier flow while their interaction with the underlying coherent Taylor rolls can be easily visualised in the $r-z$ plane as has been done in figure \ref{fig:parttraj}. For strong buoyant bubbles i.e. $Fr<1$ (c.f. figure \ref{fig:parttraj}a), the bubble primarily has an upward motion moving through different Taylor vortices as expected. For higher $Re_i$ where the turbulence is more developed and intense, the bubbles have a tendency to get trapped in the small vortical structures inside the large scale Taylor rolls. This can be observed at an axial height of $\tilde{z}\simeq0.5$ and $\tilde{z}\simeq2.5$ in the right panel of figure \ref{fig:parttraj}a. So also for larger $Fr$ number (here $Fr$=1.28 ) at fixed $Re_i$=2500 as shown in the left panel of Figure \ref{fig:parttraj}b, the bubbles have a tendency to be trapped inside the Taylor vortex (in this case at a height of $\tilde{z}\simeq3.0$) while rising along the inner cylinder wall. This behaviour has been observed also in the experiments by \citet{murai2008frictional} and \citet{fokoua2015effect}. The influence of such a motion can be seen in the contour plots of the velocity fluctuations (c.f. figure \ref{fig:azifluc}(c)). When the $Re_i$ is increased to 5000 and $Fr>1$, the smooth structured motion of the bubble inside the Taylor vortex as seen in $Re_i$=2500 is lost. Due to increased levels of turbulence in addition to loss of importance of coherent structures the bubble motion is more erratic as compared to the previous case. For both $Re_i$, a clear difference can be observed in the trajectories of the bubble depending on whether $Fr>1$ or $Fr<1$. For low $Re_i$ a bubble which smoothly passes through each Taylor vortex tends to get trapped with increase in the $Fr$ number; while for a higher $Re_i$ the bubbles which spend short periods of time in small scale vortices moves in a more erratic manner with increasing $Fr$. In all the above cases, it is to be noted that there is also a strong azimuthal motion of bubbles  

As observed in figure \ref{fig:parttraj} the bubbles dispersed into the TC system exhibit various kinds of motion such as sliding along the inner cylinder wall, outer cylinder wall, organised or erratic motion in the Taylor vortices. In order to understand and categorise such bubble motion more precisely, it would be useful to look at the mean radial and axial distribution of the bubbles and also their corresponding Reynolds numbers. For this purpose we divide the domain into three different regions (i) inner boundary layer (ibl) (ii) bulk region (bul) and (iii) outer boundary layer (obl). The method of calculating the inner and outer boundary layer thickness is described in \citet{ostilla2013optimal,ostilla2014boundary} which we discuss here in brief. A straight line is first fitted through the first three computational grid points near the inner and respective outer cylinder wall. For the bulk a straight line is  fitted through the point of inflection and the two nearest grid points. The intersection of the line drawn through the bulk and the lines near the inner and outer cylinder wall give the inner and outer boundary layer thickness, respectively.

\begin{figure}
\centerline{\includegraphics[scale=0.65]{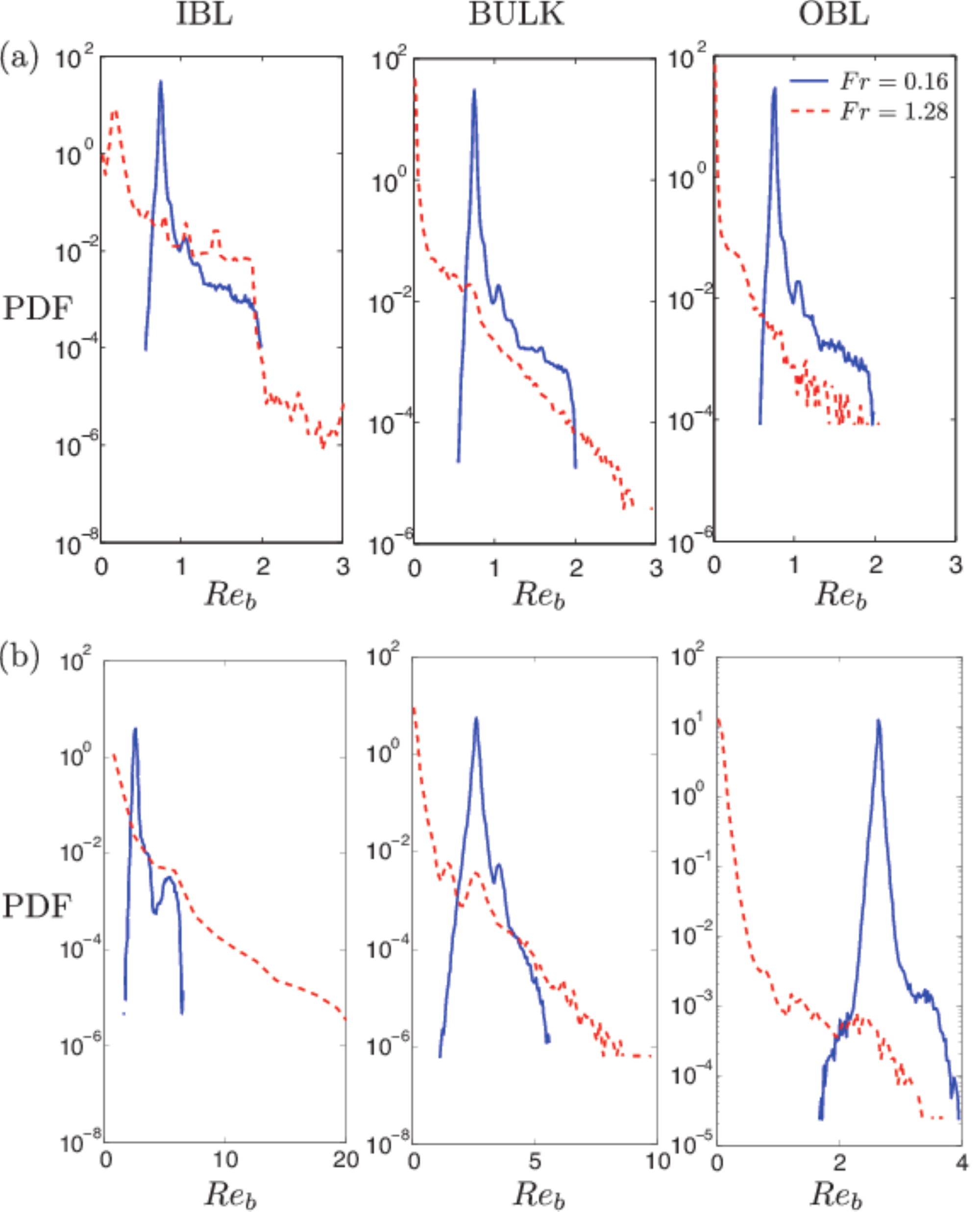}}
\caption{Probability distribution functions (pdf's) of the bubble Reynolds number for (a) $Re_i=2500$ (b) $Re_i=5000$. Left panels corresponds to the inner boundary layer region (IBL), middle panels to the bulk region (BULK), and right most panels to the outer boundary layer region (OBL).}
\label{fig:renpdf}
\end{figure}

In figure \ref{fig:renpdf}, we show normalised probability distribution functions (pdf) of the Reynolds number of the bubbles ($Re_b$) in the three different regions of the domain as described above. When the Froude number of the bubbles is less than one, $Re_b$ has a comparatively narrow distribution as compared to higher Froude number systems, regardless of the operating Reynolds number $Re_i$ of the inner cylinder and also of the position of the bubbles in the domain. When $Fr<1$, the bubbles primarily have an upward drifting motion through the Taylor rolls (c.f. figure \ref{fig:parttraj}) independent of their position in the domain which is reflected in these histograms. However for increasing inner cylinder Reynolds number $Re_i$ the pdf becomes more symmetric as compared to lower $Re_i$ which is a result of bubbles getting trapped in the intense vortical structures (also seen in the right panel of figure \ref{fig:parttraj} a). The distribution of $Re_b$ becomes wider for higher $Fr$ numbers (less buoyancy); more in the inner boundary layer than the outer boundary layer. This is a result of combination of two events; namely the ejection of plumes from the inner cylinder where the bubbles are trapped before being pulled into the Taylor roll and a relatively stronger component of the azimuthal slip velocity near the inner cylinder as compared to the outer cylinder where it is close to zero.

Now that we have looked at how the motion of bubbles depends on the $Re_i$ and $Fr$ numbers, in the next part we study the mean distribution of the bubbles in the domain. In Table \ref{tab:bubvolfr} we show the percentage volume fraction of bubbles accumulated in the inner boundary layer ($\alpha_i$), outer boundary layer ($\alpha_o$), and the bulk ($\alpha_b$). It is clear that the percentage of bubbles accumulated in the inner boundary layer is highest and for all cases it increases with $Fr$. With increasing $Fr$ the bubbles initially residing in the bulk and the outer boundary layer now migrate to the inner cylinder wall. Also the percentage distribution of the bubbles between the three different regions ($\alpha_i$, $\alpha_b$, and $\alpha_o$) does not change significantly for a fixed $Fr$ number with increasing the $Re_i$ number. 
\begin{table}
  \begin{center}
\def~{\hphantom{0}}
	\begin{tabular}{cc|c|c|c}
	$Re_i$ & $Fr$ & $\alpha_{i}$(\%) & $\alpha_{b}$(\%) & $\alpha_{o}$(\%) \\
	\hline
	\hline
	2500 & 0.16 & 0.2372 & 0.0927 & 0.0771 \\
	2500 & 1.28 & 1.1363 & 0.0346 & 0.0327 \\  
	
    5000 & 0.16 & 0.2216 & 0.0961  & 0.0675 \\
    5000 & 1.28 & 1.0822 & 0.0555  & 0.0463 \\
    \hline
    \end{tabular}
    \caption{Percentage volume fraction of bubbles in the inner boundary layer ($\alpha_{i}$), bulk ($\alpha_{b}$), and outer boundary layer ($\alpha_{o}$) for two $Re_i$ and $Fr$ numbers.}
    \label{tab:bubvolfr}
  \end{center}
\end{table}
\begin{figure}
\centerline{\includegraphics[scale=0.6]{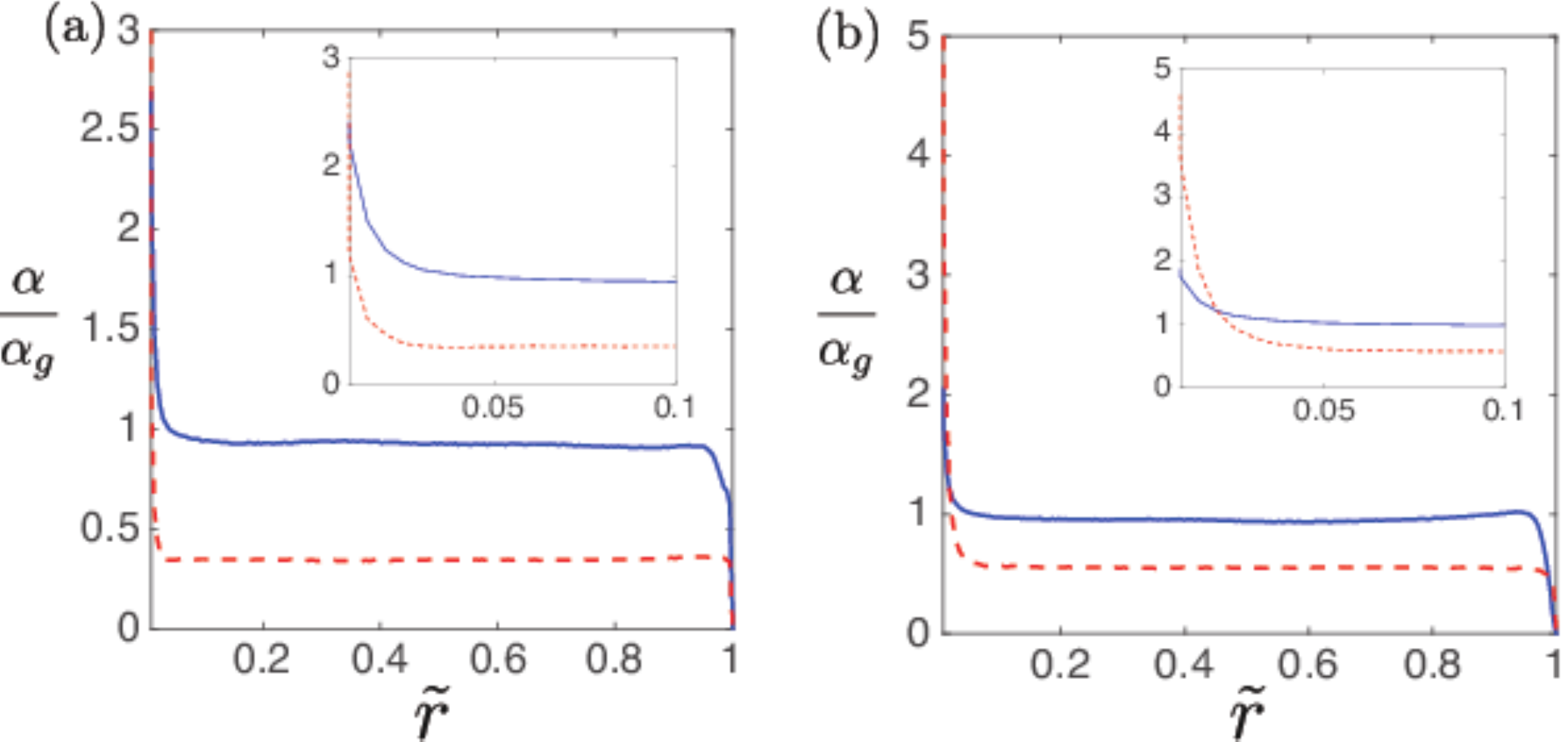}}
\caption{Radial profiles of the azimuthally, axially and time averaged local bubble volume fraction ($\alpha$) for (a) $Re_i$=2500 (b) $Re_i$=5000. Solid blue lines refer to $Fr$=0.16, while the dashed red lines refer to $Fr$=1.28. The insets show the same profile near the inner cylinder wall. $\alpha$ is normalised with $\alpha_g$=0.1\%, which is the global volume fraction in this case.}
\label{fig:radvolfr}
\end{figure}

\begin{figure}
\centerline{\includegraphics[scale=0.6]{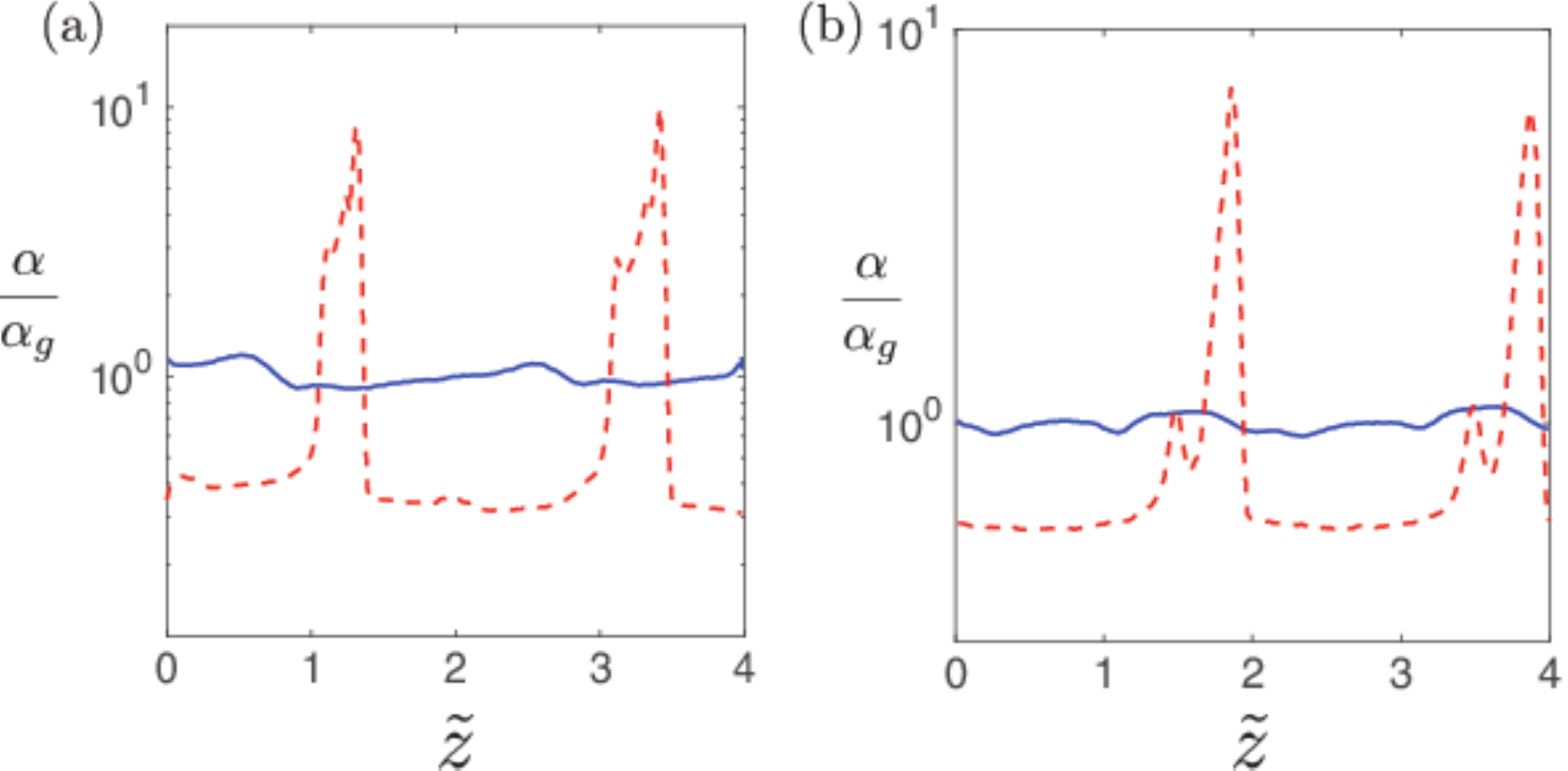}}
\caption{Axial profiles of the azimuthally, radially and time averaged local bubble volume fraction ($\alpha$) for (a) $Re_i$=2500 (b) $Re_i$=5000. Solid blue lines refer to $Fr$=0.16, while the dashed red lines refer to $Fr$=1.28. $\alpha$ is normalised with $\alpha_g$=0.1\%, which is the global volume fraction in this case. Again, the trapping of the less buoyant bubbles in the Taylor rolls becomes evident (larger Fr).}
\label{fig:axivolfr}
\end{figure}

In figure \ref{fig:radvolfr} we show the azimuthally, axially and time averaged radial profiles of the bubble volume fraction for $Re_i$=2500 and 5000 which gives a clear indication of accumulation of bubbles near the inner cylinder wall independent of the $Re_i$ number. The local volume fraction of the bubbles near the inner cylinder wall (inset of figure \ref{fig:radvolfr}) is higher for $Fr$=0.16 as compared to $Fr$=1.28. But Table \ref{tab:bubvolfr} shows that there is a higher percentage of bubble accumulation near the inner cylinder wall for $Fr$=1.28, which means that there is a higher percentage of bubbles sticking to the inner cylinder wall. When $Fr$=0.16, the local volume fraction ($\alpha$) in the bulk of the system is close to that of the global volume fraction ($\alpha_g$). With increase in $Fr$ number, the local volume fraction decreases in the bulk considerably, and more bubbles stick to the inner cylinder. These bubbles play a major role in influencing the dynamics of plume ejection and also near-wall viscous dissipation ($\hat{epsilon}$) as seen in figure \ref{fig:azifluc} and figure \ref{fig:radis2500}, respectively. While in the case of $Fr<1$, the disruption of plumes and its consequence on weakening the Taylor rolls results in low-strain rate regions in the plume impacting regions, there is no such effect when $Fr>1$ i.e. when the buoyancy of the bubbles is weaker.  

Similar to figure \ref{fig:radvolfr} we also plot the axial distribution of bubbles in the axial direction in figure \ref{fig:axivolfr}. The local volume fraction is averaged in the azimuthal, radial direction and in time. For strong buoyant bubbles ($Fr$=0.16), the axial distribution displays a largely uniform behaviour while for the increased $Fr$ number there are two distinct peaks in the local volume fraction of the bubbles for both $Re_i$ numbers. The positions of these peaks correspond to the plume ejection sites (i.e the outflow region) on the inner cylinder wall and are an effect of bubbles accumulating near these sites, before they are advected into the Taylor vortices. This distribution of bubbles when $Fr>1$ is similar to the spiral pattern of bubbles observed in the experimental results of \citet{murai2005bubble,murai2008frictional} when the Froude number is higher than one and the operating Reynolds number of the inner cylinder is $Re_i=2100-4500$.

\section{Summary}
\label{sec:summ}
The drag reduction in turbulent bubbly TC flow compared to the corresponding single phase system depends on various control parameters such as the operating Reynolds number, the diameter of the bubbles, the relative strength of buoyancy compared to the driving and the strength of coherent vortical structures. It is extremely challenging to independently control each of these parameters in experiments and study its effect on the overall drag reduction. In this work we perform numerical simulations of a two-phase TC system using an Euler-Lagrange approach which allows us to track almost 50,000 bubbles simultaneously up to an operating Reynolds number of $Re_i$=8000 and also gives us the freedom to control various parameters independently. We find that the relative strength of the buoyancy as compared to the inner cylinder driving (quantified by the Froude number $Fr$) is crucial to achieve drag reduction at higher Reynolds number. When the Froude number $Fr$ is less than one, drag reduction was observed up to $Re_i$=8000, the largest $Re_i$ we studied here. When the $Fr$ number is larger than one, almost negligible drag reduction was observed for all $Re_i$. We also observe from the averaged azimuthal velocity fields that for a fixed $Re_i$=2500, the coherent Taylor roll is much weaker for $Fr$=0.16 due to the rising strongly buoyant bubbles, as compared to $Fr$=1.28. The weakened Taylor rolls in turn imply drag reduction. The effect of the bubbles on the Taylor rolls is also observed in the contour plots of the azimuthal velocity fluctuations. By analysing the individual trajectory of the bubbles, we showed that while for a low $Re_i$=2500 the bubbles have primarily either a upward motion (lower $Fr$) or coherent motion inside the Taylor roll (larger $Fr$). Similar behaviour was observed even in the case of a higher $Re_i$, but with a more erratic movement. 

In this work we have studied numerically the effect of small bubbles ($d_b/\delta_\nu=0.55-1.25$) on the dynamics of a turbulent Taylor-Couette system using a Euler-Lagrange approach. The next line of simulations should focus on finite size bubbles which are much larger than the Kolmogorov scale. It has already been shown by experiments that almost 40 \% drag reduction can be achieved with just 4 \% volume fraction of bubbles \citep{van2013importance}. However, the Euler-Lagrange model is limited to the use of sub-Kolmogorov and Kolmogorov scale bubbles while techniques such as front tracking, immersed boundaries etc. are limited by the number of bubbles that can be simulated in a highly turbulent field. Simulation techniques which can model the deformation properties of the dispersed phase using sub-grid models are required in order to successfully advect millions of particles/droplets/bubbles in a highly turbulent carrier phase.

We would like to thank Y.Yang and E.P.van der Poel for various stimulating discussions during the development of the code. This work was supported by the Netherlands Center for Multiscale Catalytic Energy Conversion (MCEC), an NWO Gravitation programme funded by the Ministry of Education, Culture and Science of the government of the Netherlands and the FOM-CSER program. We acknowledge PRACE for awarding us access to FERMI based in Italy at CINECA under PRACE project number 2015133124 and NWO for granting us computational time on Cartesius cluster from the Dutch Supercomputing Consortium SURFsara.

\appendix
\section{}\label{app:A}

In table \ref{tab:tordets} we give the data for the Nusselt numbers used in figure \ref{fig:dr_re} and also reference data for the simulations with a larger aspect ratio ($\Gamma=8$).
\begin{table}
  \begin{center}
\def~{\hphantom{0}}
	\begin{tabular}{c| c c c c c|c c}
	$Re_i$ & Single phase & $Fr=0.16$ & $Fr=0.32$ & $Fr=1.28$ & $Fr=2.56$ & Single phase & $Fr=0.16$ \\
	       &              &           &           &           &           &    $\Gamma=8$ & $\Gamma=8$ \\ 
	\hline
	\hline
	2500 & 6.94 & 6.66 & 6.70 & 6.87 & 6.93 & 6.93 & 6.67 \\
	3500 & 7.85 & 7.57 & 7.64 & 7.83 & 7.84 & 7.86 & 7.58\\  
	5000 & 8.85 & 8.66 & 8.74 & 8.83 & 8.86 & 8.84 & 8.66\\
    8000 & 10.93 & 10.78 & 10.89 & 10.88 & 10.90 & 10.93 & 10.78\\
    \hline
    \end{tabular}
    \caption{Nusselt numbers for the single phase and two-phase cases for different $Re_i$ and $Fr$ numbers. Last two columns show the Nusselt numbers with double the aspect ratio $\Gamma=8$.}
    \label{tab:tordets}
  \end{center}
\end{table}

\bibliographystyle{jfm}

\bibliography{mylit}

\end{document}